# On Huygens' Principle, Extinction Theorem, and Equivalence Principle (Inhomogeneous Anisotropic Material System in Inhomogeneous Anisotropic Environment)

Renzun Lian

*Abstract*—Huygens' principle (HP), extinction theorem (ET), and Franz's / Franz-Harrington formulation (FHF, which is a mathematical expression of surface equivalence principle) are the important components of electromagnetic (EM) theory, and they are generalized from the following aspects.

1) Traditional HP, ET, and FHF in homogeneous isotropic environment are generalized to inhomogeneous anisotropic lossy environment. 2) Traditional FHF for homogeneous isotropic material system is generalized to inhomogeneous anisotropic lossy material system in this paper, and will be further generalized to metal-material combined system in future works. 3) The Huygens' surface in traditional HP and ET is a single closed surface. In this paper, it is generalized to the "Huygens' surface" which is constructed by multiple closed surfaces. In future works, it will be further generalized to the "Huygens' boundary" which includes some lines and open surfaces. 4) For a material body, traditional FHF has only ability to express the external scattering field and the internal total field (the summation of scattering and incident fields) in terms of the equivalent sources on material boundary, and it is generalized to formulating the internal scattering and incident fields in this paper.

In addition, the relationships among HP, ET, and FHF are studied, and it is proved that HP and ET are equivalent to each other.

*Index Terms*—Action at a distance, equivalence principle, extinction theorem (ET), Franz's formulation, Franz-Harrington formulation (FHF), Huygens' principle (HP), inhomogeneous anisotropic lossy material, material body, Green's theorem, superposition principle, the law of causality, topological additivity.

## I. Introduction

EQUIVALENCE principle is an indispensable building block for classical electromagnetic (EM) theory. The principle provides the method to express the interesting EM fields in the interesting region by using equivalent sources instead of real sources, and there are some variations of the principle, such as *volume equivalence principle (VEP)* and *surface equivalence* *principle (SEP)*. The VEP expresses the interesting EM fields in terms of equivalent volume sources, and a detailed discussion for it can be found in [1] and [2]. Compared with the VEP, the SEP is more philosophical, and has a longer history. To clarify some important concepts closely related to this paper, a brief review on the history of SEP is provided below.

The earliest researches on SEP can be dated back to C. Huygens. In 1690, he published a seminal book on the propagation of light, *Traite de la Lumiere* [3], and introduced a sophisticated principle, now known as *Huygens' principle (HP)*:

> "Each point on a primary wavefront can be considered to be a new source of a secondary spherical wave and that a secondary wavefront can be constructed as the envelope of these secondary spherical waves." [4]

Based on the principle, Huygens provided a geometrography to explain the propagation, reflection, and refraction phenomena of light, and his principle and geometrography are usually collectively referred to as *Huygens' construction*. Huygens' construction is a qualitative method instead of being quantitative, and the earliest quantitative researches on HP started with T. Young and A. Fresnel.

Around 1800, Young [5] did his famous double-slit interference experiment, and studied the diffraction phenomenon of light. To mathematically explain these new phenomena, Young introduced, in addition to the geometric-optical principle of propagation of locally-plane waves in the direction of rays, the notion of transverse transmission of the oscillation amplitudes directly along the wave-fronts [6], but his method cannot generally explain the diffraction for all cases. Until 1819, A. Fresnel [7] generally explained the diffraction by employing wave equation and boundary values, so the HP is also called as *Huygens-Fresnel principle* now. (In fact, it was shown by G. Maggi in 1888 [8] and by A. Rubinowicz in 1917 [9] and 1924 [10] that the results obtained by Fresnel's methods can be reduced by means of a mathematical transformation to the same form as predicted by Young [6], and the related theory is sometimes called as *Young-Maggi-Rubinowicz theory* [11].) However, Fresnel's original theory cannot properly describe the propagation of light in free space, because it generates the backward waves which propagate towards wave source, and then it conflicts with *the law of causality*. To suppress the backward waves, Fresnel introduced the oblique factor into his theory, so his theory is essentially a phenomenological theory.

To establish the formulation of HP on a rigorous mathematical foundation started with H. Helmholtz for steady-state (monochromatic) case in 1859 and G. Kirchhoff for time-dependent case in 1882. Now, the Helmholtz's result [12] is also known as *scalar Green's second theorem*, and the Kirchhoff's result [13], [14] is also called as *Fresnel-Kirchhoff diffraction integral formulation*. Later on, by employing the first and second kinds of half-space scalar Green's functions, Lord Rayleigh [15] and A. Sommerfeld [16] extended the Fresnel-Kirchhoff formulation to instrumental optics, and the extended formulations are now known as *Rayleigh-Sommerfeld diffraction integral formulations* [17]. Just based on the above famous works, the wave nature of light, which was hidden in *Newton's corpuscular theory* for a long period, was revealed gradually, and a very comprehensive review for this history can be found in [18] and [19].

In fact, all the above-mentioned formulations are the scalar formulations of EM waves (EM waves are essentially vectorial), so they are usually collectively referred to as *scalar diffraction formulations* [17]. The attempt to formulate *vectorial diffraction formulations* directly based on the vectorial nature of EM waves originates from the establishment of famous Maxwell's equations [20], and the earliest scholars focusing on this attempt are A. Love (1901) [21] and H. MacDonald (1911) [22]. Love introduced the concept of equivalent surface current to act as the Huygens' secondary source for the first time, and his work is now known as *Love's equivalence principle*. In 1936, S. Schelkunoff [23] extended Love's result to allow for an arbitrary EM field distribution on the both sides of a surface, and his result is now called as *Schelkunoff's equivalence principle* (Schelkunoff pointed out that this particular formulation originated from J. Larmor [24]). In 1938, J. Stratton and L. Chu [25] provided a formulation to express EM field in terms of both the normal and tangential components of field on a closed surface by employing so-called *vector Green's second theorem*, and the formulation is now called as *Stratton-Chu formulation*. In 1948, W. Franz [26] established so-called *Franz's formulation* (or called as *Kottler-Franz formulation* due to [27]), which can express EM field in terms of only tangential surface field. Later on, C.-T. Tai [28] proved that the Stratton-Chu and Franz's formulations are equivalent to each other, and pointed out that the Stratton-Chu formulation is essentially identical to the *Larmor-Tedone formulation* as described in [29]. A relatively comprehensive summarization for above vectorial diffraction formulations can be found in [2], [30], and [31].

Based on his studies on the Cauchy problem for partial differential equations, J. Hadamard [32], an outstanding mathematician, gave HP a mathematically rigorous and somewhat philosophical description, and revealed that the crucial building blocks of HP are the following three: i) *the concept of action at a distance* (this concept originates from M. Faraday and J. Maxwell [20]), ii) *the law of causality*, and iii) *the principle of superposition*. The first block implies that the formulation of HP needs to employ the field propagator (Green's function), and the second block implies that the propagator should prop-agate away from source rather than being towards source (i.e., the propagator should satisfy *Sommerfeld's radiation condition* [33]), and the third block implies that the formulation of HP will be expressed as integral. Hence, all formulations mentioned above are the integral formulations basing on outgoing Green's functions.

Obviously, the establishment for HP and its mathematical formulation experienced an evolution from qualitative to quantitative and from scalar to vectorial. In classical electromagnetics framework, the Franz's formulation reaches the peak of the evolution as stated by Prof. Tai that:

> "It seems obvious that the Franz formula is conceptually simpler since it requires only the tangential components of the field on the closed surface, while the Stratton-Chu formula requires the normal components as well. Most important of all, when the field has an edge singularity on the surface of integration the Larmor-Tedone formula or Stratton-Chu formula must be modified as shown by Kottler in order to make the resultant field Maxwellian." [28]

Because of this, the Franz's formulation has been widely applied in EM engineering society. For example, the famous PMCHWT-based scattering integral equation (A. Poggio and E. Miller [34], Y. Chang and R. Harrington [35], W. Wu [36], and Tsai) and the PMCHWT-based characteristic mode (CM) formulation [35] for homogeneous isotropic material bodies were established basing on the Franz's formulation.

The formulation utilized by Chang and Harrington in [35] is essentially the Franz's formulation, but the former is more advantageous than the latter in the following aspects. a) The former is more concise than the latter in mathematical form. b) The former expresses various fields in terms of an identical set of equivalent currents, i.e., the external scattering field and the internal total field of material body are simultaneously expressed as the functions of the equivalent surface currents defined by using boundary tangential total fields. Based on these features, the former is more popular than the latter in computational electromagnetics and EM engineering, though the former is essentially the latter from the perspective of EM theory. In addition, Chang and Harrington provided their formulation in [35] by directly citing Harrington's classical book [31], so their formulation is particularly called as *Franz-Harrington formulation (FHF)* in this paper.

Although FHF has had many successful applications as mentioned above, it also has some limitations, and this paper does some works to remove the limitations.

**The limitations on EM media**

The traditional HP and ET are only valid for homogeneous isotropic environment, and the traditional FHF is only valid for a homogeneous isotropic material body in homogeneous isotropic environment. It is still an unsolved problem how to establish the HP, ET, and FHF of inhomogeneous anisotropic lossy material body in inhomogeneous anisotropic lossy environment, and it is done in this paper.

In the appendixes of this paper, the EM field in inhomogeneous anisotropic lossy environment is expressed in terms of its tangential components on a closed surface, based on *general-* *ized vector-dyadic Green's second theorem*. The reason to utilize the vector-dyadic version of Green's second theorem instead of the vector version used in [25] is based on Prof. Tai's observation:

> "… the most compact formulation appears to be the one based on the dyadic Green's function pertaining to the vector wave equation for $\vec{E}$ and $\vec{H}$ …" [28]

The reason to utilize the generalized version of vector-dyadic Green's second theorem instead of the traditional version used in [37] is that the traditional one is suitable for neither inhomogeneous media nor anisotropic media. Based on the results given in appendixes, the traditional HP, ET, and FHF are generalized in Secs. II-IV from the following aspects.

• Based on the integral formulations given in appendixes, the traditional HP and ET are generalized to inhomogeneous anisotropic lossy environment, and the traditional FHF is generalized to an inhomogeneous anisotropic lossy body in inhomogeneous anisotropic lossy environment, in Secs. II and III.

• Based on the results obtained in Secs. II and III, the FHF is further generalized to the piecewise inhomogeneous anisotropic lossy body in inhomogeneous anisotropic lossy environment, in Sec. IV. The adjective "piecewise" means that the material parameters are discontinuous on two sides of the interface between two different media, as shown in Fig. 1 (d).

**The limitations on topological structure**

The Huygens' surface used in traditional HP and ET is a single closed surface. The traditional FHF is only valid for the EM system $V_{sys}^{mat}$ constructed by a simply connected material body $V_{sim}^{mat}$, i.e., $V_{sys}^{mat} = V_{sim}^{mat}$. In this paper, the traditional HP, ET, and FHF are generalized from the following aspects.

• In Secs. III and IV, the Huygens' surface is generalized to multiple closed surfaces.

• In Sec. III, the EM system $V_{sys}^{mat}$ constructed by a multiply connected material body $V_{mul}^{mat}$ is considered, i.e., $V_{sys}^{mat} = V_{mul}^{mat}$, and the FHF for $V_{mul}^{mat}$ is derived.

• The systems focused on by Secs. II and III are connected, and the results corresponding to these connected systems are further generalized to the non-connected system in Sec. IV.

The above terminologies "connected, non-connected, simply connected, and multiply connected" are commonly used terms in point set topology, and their rigorous definitions can be found in [38]. The "non-connected" can be vividly understood as that there exist some different parts of system, such that the parts don't contact with each other, as shown in Fig. 1 (c); if the system is not non-connected, it is connected, as shown in Fig. 1 (a); the "simply connected" can be vividly understood as that there doesn't exist any hole on material body, as shown in Fig. 1 (a); the "multiply connected" can be vividly understood as that there exist some holes on material body, as shown in Fig. 1 (b).

**The limitations on expressing EM field**

When a simply connected material body $V_{sim}^{mat}$ is considered, the whole three-dimensional Euclidean space $\mathbb{R}^3$ is divided into two parts (the interior of $V_{sim}^{mat}$ and the exterior of $V_{sim}^{mat}$) by the material boundary $\partial V_{sim}^{mat}$, as shown in Fig. 2. The traditional FHF [35] has only ability to express the external scattering field $\vec{F}_+^{sca}$ and internal total field $\vec{F}_-^{tot}$ (the summation of internal incident field $\vec{F}_-^{inc}$ and internal scattering field $\vec{F}_-^{sca}$) in terms of an identical set of equivalent surface currents on $\partial V_{sim}^{mat}$. In this paper, the traditional FHF is generalized from the following aspects.

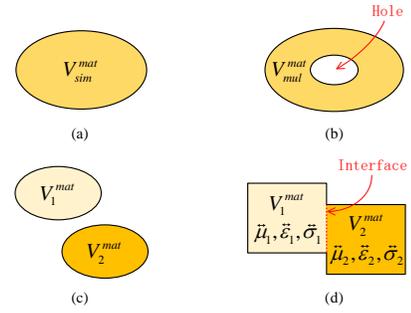

Fig. 1. (a) A simply connected material body; (b) a multiply connected material body; (c) a non-connected system constructed by two material bodies; (d) piecewise inhomogeneous anisotropic lossy material body.

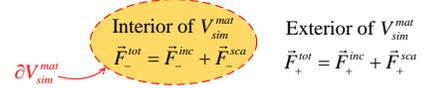

Fig. 2. Whole space is divided into two parts by the boundary of a simply connected material body.

• In Sec. II-A, the FHF for $\vec{F}_-^{inc}$ and $\vec{F}_-^{sca}$ are derived, and they are valuable for material CM theory as discussed in Sec. II-D and as exhibited in Sec. V.

In addition, this paper also does some works in the following aspects.

**On mathematically formulating HP**

The traditional FHF of $\vec{F}_+^{sca}$ is usually viewed as the mathematical expression of scattering field HP, but it will generate backward waves in the interior of material body, so it conflicts with the law of causality. In fact, the backward wave problem also exists in traditional scalar diffraction theory as pointed out by D. Miller in [39]. In addition, the traditional FHF of $\vec{F}_-^{tot}$ is sometimes classified into the *extinction theorem (ET)* family (ET is also known as *Ewald-Oseen extinction theorem* [40], [41], due to the works of C. Oseen [42] *et al.*). However, the FHF of $\vec{F}_-^{tot}$ cannot guarantee the null result in whole exterior of material body, though it indeed generates null tangential field on the external surface of material boundary.

To clarify the reasons leading to above problems, the relationships among HP, ET, FHF, and SEP are carefully studied in Sec. II, and it is found out that:

• The mathematical formulation of HP is equivalent to ET.

• The FHF of $\vec{F}_-^{tot}$ satisfies so-called *weak extinction theorem*, and it should not be classified into ET family.

• HP is a special SEP, but SEP is not necessarily HP.

• FHF is not the mathematical expression of HP, and it is only the mathematical expression of SEP.

**On topological additivity**

The EM systems $V_{sys}^{mat}$ considered in Secs. II, III, and IV have different topological structures, i.e., $V_{sys}^{mat} = V_{sim}^{mat}$ in Sec. II, and $V_{sys}^{mat} = V_{mul}^{mat}$ in Sec. III, and $V_{sys}^{mat} = V_{sim}^{mat} \bigcup V_{mul}^{mat}$ in Sec. IV. It is pointed out in Sec. IV-B that the mathematical formulation of HP and ET satisfy so-called *topological additivity*, i.e., the HP/ET of whole EM system equals to the summation of the HP/ET corresponding to all sub-systems, and this property is consistent with the principle of superposition. Then, the FHF of $\vec{F}_-^{inc}$ and $\vec{F}_+^{sca}$ also satisfy topological additivity, because they are essentially the summation of incident field HP and scattering field HP as pointed out in Sec. II-C. However, the FHF of $\vec{F}^{tot}$ and $\vec{F}^{sca}$ don't satisfy topological additivity.

• To guarantee the topological additivity for the FHF of $\vec{F}^{tot}$ and $\vec{F}^{sca}$, a so-called *piecewise Green's function* is proposed in Sec. IV-B.

For the convenience of the following discussions, the symbolic system of this paper is summarized here. The $e^{j\omega t}$ convention is used in this paper. The conductivity, permittivity, and permeability of material system are denoted as $\ddot{\sigma}(\vec{r})$, $\ddot{\varepsilon}(\vec{r})$, and $\ddot{\mu}(\vec{r})$ respectively, and the conductivity, permittivity, and permeability of environment are denoted as $\ddot{\sigma}_{env}(\vec{r})$, $\ddot{\varepsilon}_{env}(\vec{r})$, and $\ddot{\mu}_{env}(\vec{r})$ respectively, and all these parameters are restricted to being symmetrical two-order tensors, because many commonly used anisotropic materials (such as crystal) have symmetrical material parameters [19], [43], [44]. Some concepts related to point set topology (such as the open set $\Omega$, boundary $\partial\Omega$, closure $\text{cl}\Omega$, interior $\text{int}\Omega$, and exterior $\text{ext}\Omega$) need to be utilized, and the rigorous mathematical definitions for the first four can be found in [38], and the last one is defined as that $\text{ext}\Omega \triangleq \mathbb{R}^3 \setminus \text{cl}\Omega$. Obviously, both the $\text{int}\Omega$ and $\text{ext}\Omega$ are open sets [38]. When an external excitation $\vec{F}^{inc}$ incidents on EM system $V_{sys}^{mat}$ ($V_{sys}^{mat} = V_{sim}^{mat} \cup V_{mul}^{mat}$), the scattering sources will be excited on simply connected body $V_{sim}^{mat}$ and multiply connected body $V_{mul}^{mat}$, and then the scattering fields $\vec{F}_{sim}^{sca}$ and $\vec{F}_{mul}^{sca}$ are generated by $V_{sim}^{mat}$ and $V_{mul}^{mat}$ respectively. The summation of $\vec{F}_{sim}^{sca}$ and $\vec{F}_{mul}^{sca}$ is just the total scattering field $\vec{F}^{sca}$ generated by $V_{sys}^{mat}$, i.e., $\vec{F}^{sca} = \vec{F}_{sim}^{sca} + \vec{F}_{mul}^{sca}$. The summation of $\vec{F}^{inc}$ and $\vec{F}^{sca}$ is total field $\vec{F}^{tot}$, i.e., $\vec{F}^{tot} = \vec{F}^{inc} + \vec{F}^{sca} = \vec{F}^{inc} + \vec{F}_{sim}^{sca} + \vec{F}_{mul}^{sca}$, as shown in Fig. 3.

In addition, we sincerely wish that the appendixes are read before reading the main body of this paper.

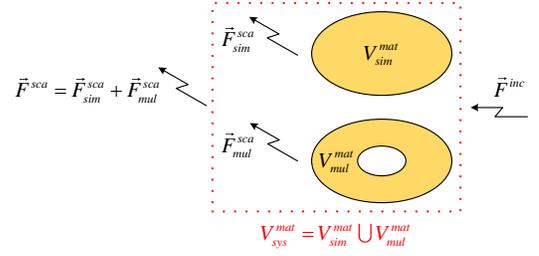

Fig. 3. An external excitation field incidents on the material system which is constructed by a simply connected body and a multiply connected body.

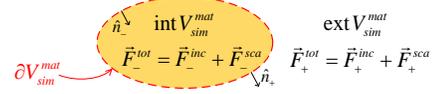

Fig. 4. Various domains and fields related to a simply connected material body.

## II. A SINGLE SIMPLY CONNECTED INHOMOGENEOUS ANISOTROPIC LOSSY MATERIAL BODY

The EM system $V_{sys}^{mat}$ focused on by this section is a simply connected inhomogeneous anisotropic lossy body $V_{sim}^{mat}$, i.e., $V_{mul}^{mat} = \varnothing$ and $V_{sys}^{mat} = V_{sim}^{mat}$, as shown in Fig. 4. Hence, $\vec{F}_{mul}^{sca} = 0$, and $\vec{F}^{sca} = \vec{F}_{sim}^{sca}$, and $\vec{F}^{tot} = \vec{F}^{inc} + \vec{F}^{sca} = \vec{F}^{inc} + \vec{F}_{sim}^{sca}$. Based on the results given in appendixes, the traditional HP, ET, and FHF of a simply connected homogeneous isotropic material body in homogeneous isotropic environment are generalized to inhomogeneous anisotropic lossy case, in this section.

### A. Huygens' principle and extinction theorem

The integral formulations (C-5) and (C-7) given in Appendix C can be uniformly written as follows:

$$\left.\begin{array}{l}(\vec{r} \in \text{ext}V_{sim}^{mat}) \quad , \quad 0 \\ (\vec{r} \in \text{int}V_{sim}^{mat}) \quad , \quad \vec{F}^{inc}(\vec{r})\end{array}\right\} = \oiint_{\partial V_{sim}^{mat}} \ddot{G}_{env}^{JF}(\vec{r},\vec{r}') \cdot [\hat{n}_- \times \vec{H}^{inc}(\vec{r}')]dS' + \oiint_{\partial V_{sim}^{mat}} \ddot{G}_{env}^{MF}(\vec{r},\vec{r}') \cdot [\vec{E}^{inc}(\vec{r}') \times \hat{n}_-]dS' \quad (1)$$

where $F = E, H$, and $\hat{n}_-$ is the inward normal vector of $\partial V_{sim}^{mat}$. The $\ddot{G}_{env}^{JF}(\vec{r},\vec{r}')$ and $\ddot{G}_{env}^{MF}(\vec{r},\vec{r}')$ in (1) are the environment dyadic Green's functions. Following the manner to express convolution integrals in [45], (1) is rewritten as the following (1') to compact the integral formulation, and the other convolution integrals appeared in this paper will be similarly expressed.

$$\left.\begin{array}{l}\text{ext}V_{sim}^{mat} : \quad 0 \\ \text{int}V_{sim}^{mat} : \quad \vec{F}^{inc}\end{array}\right\} = [\ddot{G}_{env}^{JF} * (\hat{n}_- \times \vec{H}^{inc})]_{\partial V_{sim}^{mat}} + [\ddot{G}_{env}^{MF} * (\vec{E}^{inc} \times \hat{n}_-)]_{\partial V_{sim}^{mat}} = [\ddot{G}_{env}^{JF} * (\hat{n}_- \times \vec{H}^{inc}) + \ddot{G}_{env}^{MF} * (\vec{E}^{inc} \times \hat{n}_-)]_{\partial V_{sim}^{mat}} \quad (1')$$

where $F = E, H$. The integral formulations (C-9) and (C-10) given in Appendix C can be uniformly written as follows:

$$\left.\begin{array}{l}\text{ext}V_{sim}^{mat} : \quad \vec{F}^{sca} \\ \text{int}V_{sim}^{mat} : \quad 0\end{array}\right\} = [\ddot{G}_{env}^{JF} * (\hat{n}_+ \times \vec{H}^{sca}) + \ddot{G}_{env}^{MF} * (\vec{E}^{sca} \times \hat{n}_+)]_{\partial V_{sim}^{mat}} \quad (2)$$

where $F = E, H$. In (2), $\hat{n}_+$ is the outward normal vector of $\partial V_{sim}^{mat}$, and $\hat{n}_+ = -\hat{n}_-$ on whole $\partial V_{sim}^{mat}$.

(1') and (2) imply that the equivalent secondary sources $\{\hat{n}_- \times \vec{H}^{inc}, \vec{E}^{inc} \times \hat{n}_-\}$ and $\{\hat{n}_+ \times \vec{H}^{sca}, \vec{E}^{sca} \times \hat{n}_+\}$ will establish the zero fields on whole $\text{ext}V_{sim}^{mat}$ and whole $\text{int}V_{sim}^{mat}$ respectively, i.e., they will not generate any backward wave, so they are usually called as extinction theorem. In fact, (1') and (2) are just the mathematical formulations of HP corresponding to incident field $\vec{F}^{inc}$ and scattering field $\vec{F}^{sca}$ respectively.

### B. Generalized Franz-Harrington formulation and weak extinction theorem

Because of that $\vec{F}^{tot} = \vec{F}^{inc} + \vec{F}^{sca}$, the difference between (1') and (2) is as follows:

$$\left.\begin{array}{l}\text{ext}V_{sim}^{mat} : \quad -\vec{F}^{sca} \\ \text{int}V_{sim}^{mat} : \quad \vec{F}^{inc}\end{array}\right\} = [\ddot{G}_{env}^{JF} * \vec{J}^{ES} + \ddot{G}_{env}^{MF} * \vec{M}^{ES}]_{\partial V_{sim}^{mat}} \quad (3)$$

where the $\vec{J}^{ES}$ and $\vec{M}^{ES}$ are defined as follows:

$$\vec{J}^{ES}(\vec{r}) \triangleq \hat{n}_-(\vec{r}) \times [\vec{H}^{tot}(\vec{r}')]_{\vec{r}' \to \vec{r}} \quad , \quad (\vec{r} \in \partial V_{sim}^{mat}) \quad (4.1)$$

$$\vec{M}^{ES}(\vec{r}) \triangleq \left[\vec{E}^{tot}(\vec{r}')\right]_{\vec{r}'\to\vec{r}} \times \hat{n}_-(\vec{r}) \;,\quad (\vec{r} \in \partial V_{sim}^{mat}) \quad (4.2)$$

in which $\vec{r}' \in \text{int}\, V_{sim}^{mat}$, and the superscript "$ES$" is the acronyms of term "equivalent surface". Based on (4.1) and (4.2), (C-11) can be rewritten as follows:

$$\text{int}\, V_{sim}^{mat}\;:\;\vec{F}^{tot} = \left[\vec{\vec{G}}_{sim}^{JF} * \vec{J}^{ES} + \vec{\vec{G}}_{sim}^{MF} * \vec{M}^{ES}\right]_{\partial V_{sim}^{mat}} \quad (5)$$

where the $\vec{\vec{G}}_{sim}^{JF}(\vec{r},\vec{r}')$ and $\vec{\vec{G}}_{sim}^{MF}(\vec{r},\vec{r}')$ are the dyadic Green's functions corresponding to the inhomogeneous anisotropic lossy material body $V_{sim}^{mat}$. Based on (3) and (5) and that $\vec{F}^{sca} = \vec{F}^{tot} - \vec{F}^{inc}$, the following formulation for the $\vec{F}^{sca}$ on $\text{int}\, V_{sim}^{mat}$ can be obtained:

$$\text{int}\, V_{sim}^{mat}\;:\;\vec{F}^{sca} = \left[\Delta\vec{\vec{G}}_{sim}^{JF} * \vec{J}^{ES} + \Delta\vec{\vec{G}}_{sim}^{MF} * \vec{M}^{ES}\right]_{\partial V_{sim}^{mat}} \quad (6)$$

where

$$\Delta\vec{\vec{G}}_{sim}^{JF}(\vec{r},\vec{r}') \triangleq \vec{\vec{G}}_{sim}^{JF}(\vec{r},\vec{r}') - \vec{\vec{G}}_{env}^{JF}(\vec{r},\vec{r}') \quad (7.1)$$

$$\Delta\vec{\vec{G}}_{sim}^{MF}(\vec{r},\vec{r}') \triangleq \vec{\vec{G}}_{sim}^{MF}(\vec{r},\vec{r}') - \vec{\vec{G}}_{env}^{MF}(\vec{r},\vec{r}') \quad (7.2)$$

for any $\vec{r},\vec{r}' \in \text{int}\, V_{sim}^{mat}$.

In (3), (5), and (6), the internal incident field $\vec{F}_-^{inc}$, internal scattering field $\vec{F}_-^{sca}$, internal total field $\vec{F}_-^{tot}$, and external scattering field $\vec{F}_+^{sca}$ are simultaneously expressed in terms of an identical set of equivalent surface currents $\{\vec{J}^{ES}, \vec{M}^{ES}\}$. (3), (5), and (6) are collectively referred to as *generalized Franz-Harrington formulation (GFHF)* of a simply connected inhomogeneous anisotropic lossy material body in inhomogeneous anisotropic lossy environment, and the reason to utilize adjective "generalized" is that the traditional FHF has only ability to express the $\vec{F}_+^{sca}$ and $\vec{F}^{tot}$ corresponding to a simply connected homogeneous isotropic material body in homogeneous isotropic environment. In addition, the above GFHF for a simply connected body will be further generalized to multiply connected case in Sec. III and to non-connected case in Sec. IV.

Sometimes, (5) is also called as extinction theorem just like calling (1') and (2). However, it should be emphasized that (5) cannot establish null field in whole region $\text{ext}\, V_{sim}^{mat}$, though it indeed can establish null tangential field on the external surface of $\partial V_{sim}^{mat}$. The reason leading to this will be carefully discussed in the following Sec. II-C. Based on this observation, (5) is particularly called as *weak extinction theorem* to be distinguished from the extinction theorems (1') and (2).

*C. Relationships among Huygens' principle, extinction theorem, Franz-Harrington formulation, and surface equivalence principle*

Based on Hadamard's work [32], the HP can be divided in the form of a syllogism as follows.

Major Premise: The action of phenomena produced at the instant $t=0$ on the state of matter at the latter time $t=t_0$ takes place by the mediation of every intermediary instant $t=t'$, ... (here, $0 < t' < t_0$).

Minor Premise: If we produce a luminous disturbance localized in a neighborhood of $\vec{r}=0$, its effect after an elapsed time $t_0$ will be localized in a neighborhood of the spherical surface $|\vec{r}|=ct_0$.

Conclusion: In order to calculate the effect of our initial luminous phenomenon produced at $\vec{r}=0$ at $t=0$, we may replace it by a proper system of disturbances taking place at $t=t'$ and distributed over the spherical surface $|\vec{r}|=ct_0$.

In fact, Hadamard's major premise is essentially *the concept of action at a distance*, i.e., the EM interaction is implemented by propagation; Hadamard's minor premise is essentially *the law of causality*, i.e., the propagation of EM field should be away from real source instead of being towards real source; Hadamard's conclusion is essentially the Huygens' construction based on two premises and *the principle of superposition*.

**The relationships between HP and ET**

Obviously, the law of causality implies that the mathematical formulation of HP must establish null field in the backward direction of Huygens' surface, so the mathematical formulation of HP must satisfy ET, such as the incident field HP (1') and the scattering field HP (2).

It will be proved as below that: *an extinction-type formulation corresponds to the HP of a field*. Let us consider the following extinction-type convolution integral:

$$\left.\begin{array}{l}\Omega\;:\;\vec{F}\\ \mathbb{R}^3\setminus\text{cl}\,\Omega\;:\;0\end{array}\right\} = \left[\vec{\vec{G}}_{env}^{JF} * (\hat{n}_{\to\Omega}\times\vec{H}) + \vec{\vec{G}}_{env}^{MF} * (\vec{E}\times\hat{n}_{\to\Omega})\right]_{\partial\Omega\setminus S_\infty} \quad (8)$$

where $F=E,H$. In (8), $\partial\Omega$ is the boundary surface of open domain $\Omega$, if $\Omega$ is a finite domain; $\partial\Omega\cup S_\infty$ is the boundary surface of open domain $\Omega$, if $\Omega$ is an infinite domain. $S_\infty$ is a spherical surface at infinity. $\partial\Omega\setminus S_\infty$ means that the integral domain of (8) is a limited closed surface, and then the field $\vec{F}$ and Green's functions $\vec{\vec{G}}_{env}^{JF}$ and $\vec{\vec{G}}_{env}^{MF}$ in (8) must satisfy Sommerfeld's radiation condition, or the surface $S_\infty$ cannot be excluded from integral domain [40]. $\hat{n}_{\to\Omega}$ is the normal vector of integral surface, and points to domain $\Omega$. Due to the radiation condition of field $\vec{F}$, it can be concluded that the real sources $\{\vec{J},\vec{M}\}$ of $\vec{F}$ distribute in a limited region denoted as $V$, and then the $\vec{F}$ has the following integral formulation:

$$\mathbb{R}^3\;:\;\vec{F} = \left[\vec{\vec{G}}_{env}^{JF} * \vec{J} + \vec{\vec{G}}_{env}^{MF} * \vec{M}\right]_V \quad (9)$$

based on a similar method to deriving (C-4) from (C-1).

If the Green's functions used in (8) and (9) satisfy the following Maxwell's equations:

$$\nabla\times\vec{\vec{G}}_{env}^{JH}(\vec{r},\vec{r}') = \vec{I}\delta(\vec{r}-\vec{r}') + j\omega\vec{\vec{\varepsilon}}_{env;c}(\vec{r})\cdot\vec{\vec{G}}_{env}^{JE}(\vec{r},\vec{r}')$$
$$\nabla\times\vec{\vec{G}}_{env}^{JE}(\vec{r},\vec{r}') = -j\omega\vec{\vec{\mu}}_{env}(\vec{r})\cdot\vec{\vec{G}}_{env}^{JH}(\vec{r},\vec{r}') \quad (10.1)$$

and

$$\nabla\times\vec{\vec{G}}_{env}^{MH}(\vec{r},\vec{r}') = j\omega\vec{\vec{\varepsilon}}_{env;c}(\vec{r})\cdot\vec{\vec{G}}_{env}^{ME}(\vec{r},\vec{r}')$$
$$\nabla\times\vec{\vec{G}}_{env}^{ME}(\vec{r},\vec{r}') = -\vec{I}\delta(\vec{r}-\vec{r}') - j\omega\vec{\vec{\mu}}_{env}(\vec{r})\cdot\vec{\vec{G}}_{env}^{MH}(\vec{r},\vec{r}') \quad (10.2)$$

then (8) implies that the field $\vec{F}$ satisfies the following ho-mogeneous Maxwell's equations in region $\Omega$:

$$\begin{aligned}\nabla \times \vec{H}(\vec{r}) &= j\omega\vec{\vec{\varepsilon}}_{env;c}(\vec{r}) \cdot \vec{E}(\vec{r}) \\ \nabla \times \vec{E}(\vec{r}) &= -j\omega\vec{\vec{\mu}}_{env}(\vec{r}) \cdot \vec{H}(\vec{r})\end{aligned}, \quad (\vec{r} \in \Omega) \quad (11)$$

i.e., the region $\Omega$ is source-free, where the derivation of (11) from (8) is completely similar to the inverse process of deriving (C-5) from (C-1). The (11) implies that the source $\{\vec{J},\vec{M}\}$ must distribute in region $\mathbb{R}^3 \setminus \Omega$, i.e.,

$$\begin{aligned}\nabla \times \vec{H}(\vec{r}) &= \vec{J}(\vec{r}) + j\omega\vec{\vec{\varepsilon}}_{env;c}(\vec{r}) \cdot \vec{E}(\vec{r}) \\ \nabla \times \vec{E}(\vec{r}) &= -\vec{M}(\vec{r}) - j\omega\vec{\vec{\mu}}_{env}(\vec{r}) \cdot \vec{H}(\vec{r})\end{aligned}, \quad (\vec{r} \in \mathbb{R}^3 \setminus \Omega). \quad (12)$$

The above (11) and (12) imply that the $\partial\Omega$ encloses all sources $\{\vec{J},\vec{M}\}$. $S_\infty$ is always excluded from the integral surface in (8), even if $S_\infty \subset \partial\Omega$, so it can be viewed as that the integral surface of (8) encloses all sources $\{\vec{J},\vec{M}\}$.

Just like deriving (1') from (C-1), the following ET can be derived from (12):

$$\left.\begin{aligned}V_i^{interesting} &: \vec{F} \\ V_i^{excluded} &: 0\end{aligned}\right\} = \left[\vec{\vec{G}}_{env}^{JF} * \left(\hat{n}_{\to V_i^{interesting}} \times \vec{H}\right) + \vec{\vec{G}}_{env}^{MF} * \left(\vec{E} \times \hat{n}_{\to V_i^{interesting}}\right)\right]_{S_i}$$
(13)

for any $i = 1, 2, \cdots$. In (13), $S_i$ is a closed surface which encloses whole source region $\mathbb{R}^3 \setminus \Omega$, and $S_{i+1}$ encloses $S_i$ for any $i = 1, 2, \cdots$, and $V_i^{interesting}$ and $V_i^{excluded}$ are respectively the interesting and excluded regions restricted by surface $S_i$, and the case that the source region $\mathbb{R}^3 \setminus \Omega$ is a limited region is shown in Fig. 5. In fact, this (13) is just the mathematical expression of Hadamard's syllogism, i.e., the EM field generated by a system is the superposition of the fields generated by all sub-sources (*the principle of superposition*), and the field will propagate (*the concept of action at a distance*) outward rather than being inward (*the law of causality*). Then, the extinction-type formulation (8) must correspond to the HP of a field.

Based on the above observations, it can be concluded that HP and ET are essentially equivalent to each other, i.e., the mathematical formulation of the HP for any field satisfies ET, and any ET (8) corresponds to the HP of a field.

**The necessary conditions to establish an extinction-type formulation**

Based on the process to derive (13) from (8), it is easy to find out that the necessary conditions (NCs) to establish an extinction-type formulation are as follows:

- NC-1, On real source of field: The (9) implies that *the real source of interesting field must distribute in a limited region*.
- NC-2, On Huygens' surface: The (11) and (12) imply that *the Huygens' surface must enclose all real source of interesting field*. As exhibited in (13), the reasonable Huygens' surface is not unique, and the boundary of source region is a natural and the smallest one.
- NC-3.1, On interesting and excluded regions: Whole space is divided into two parts by Huygens' surface as illustrated in (13). *The interesting region must be the source-free one, and the other one is the excluded region*, as illustrated in formulations (11)-(13) and Fig. 5.
- NC-3.2, On interesting field: *The interesting field must satisfy a homogeneous Maxwell's equations in interesting region*, as illustrated in (11). At the same time, *the interesting field must satisfy an inhomogeneous Maxwell's equations in excluded region*, as illustrated in (12).
- NC-4, On Huygens' secondary source: *The Huygens' secondary source on Huygens' surface must correspond to the interesting field*, as illustrated in (8) and (13).
- NC-5, On propagator / Green's function: *The Maxwell's equations of propagators must have the same material parameters as the ones which are satisfied by interesting field*, as illustrated in (10). *The propagators must satisfy the Sommerfeld's radiation condition*, as concluded below the formulation (8).

In fact, these conditions are also the necessary conditions to mathematically formulate HP, because the mathematical formulation of HP must satisfy ET as concluded in the previous part of this subsection. In the following parts of this subsection, the relationships between HP and FHF will be clarified based on above conditions.

**The relationships between HP and FHF**

Based on above discussions in this section, it can be concluded that:

A) The FHF (3) is neither the mathematical formulation of HP corresponding to $\vec{F}^{inc}$ nor the mathematical formulation of HP corresponding to $\vec{F}^{sca}$, because it doesn't satisfy the condition NC-4, i.e., the equivalent currents used in (3) correspond to neither $\vec{F}^{inc}$ nor $\vec{F}^{sca}$. In fact, the FHF (3) is solely the difference between the incident field HP (1') and the scattering field HP (2), i.e.,

$$\text{Formulation (3)} = \text{Formulation (1')} - \text{Formulation (2)}. \quad (14)$$

This is just the reason why (3) will generate some backward waves, i.e., why (3) conflicts with the law of causality.

B) The FHF (5) is not the mathematical formulation of HP corresponding to internal total field $\vec{F}_-^{tot}$, because it doesn't satisfy the condition NC-2. Specifically, the real source of $\vec{F}^{tot}$ includes both the $\{\vec{J}^{inc}, \vec{M}^{inc}\}$ generating $\vec{F}^{inc}$ and the $\{\vec{J}^{SV}, \vec{M}^{SV}\}$ generating $\vec{F}^{sca}$, but the surface $\partial V_{sim}^{mat}$ doesn't enclose all these sources. To formulate the HP corresponding to $\vec{F}^{tot}$, the Huygens' surface should enclose both $\{\vec{J}^{inc}, \vec{M}^{inc}\}$ and $\{\vec{J}^{SV}, \vec{M}^{SV}\}$.

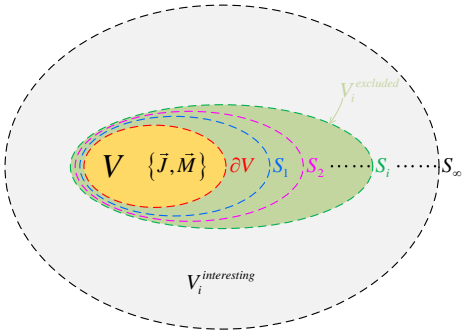

Fig. 5. The diagram of formulation (13).

C) The FHF (6) is not the mathematical formulation of HP corresponding to internal scattering field $\vec{F}^{sca}_-$, because it doesn't satisfy the conditions NC-3 and NC-4. In fact, (6) is solely the difference between (5) and (3), i.e.,

$$\text{Formulation (6)} = \text{Formulation (5)} - \text{Formulation (3)}. \quad (15)$$

**Summary**

Because of the above observations, the relationships among HP, ET, FHF, and SEP are illustrated in Fig. 6. From Fig. 6, it can be concluded that:
- HP and ET are equivalent to each other.
- HP is a special SEP, but SEP is not necessarily HP. HP can be particularly called as *physical equivalence principle*, because it simultaneously satisfies the concept of action at a distance, the law of causality, and the principle of superposition.
- FHF is only the mathematical expression of SEP instead of the mathematical expression of HP, so the surface currents (4) used in FHF should be called as *equivalent surface currents*, but should not be viewed as *Huygens' secondary sources*.
- Compared with the incident field HP (1') and the scattering field HP (2), the values of FHF are mainly manifested in that various fields are uniformly expressed in terms of an identical set of currents $\{\vec{J}^{ES}, \vec{M}^{ES}\}$, and this feature is very valuable for many engineering applications as pointed out in the following Sec. II-D and as exhibited in Sec. V.

### D. Applications of generalized Franz-Harrington formulation

In this subsection, some typical engineering applications related to GFHF are simply mentioned.

**Application on solving EM scattering problem**

The well-known PMCHWT equation [34]-[36] is derived from FHF, and it is widely applied to solving EM scattering problem, but it is only suitable for a simply connected homogeneous isotropic body in homogeneous isotropic environment. Obviously, the traditional PMCHWT equation and its related applications can be easily generalized to the inhomogeneous anisotropic lossy case, by employing GFHF. In computational electromagnetics, the traditional FHF is usually written as the operator forms based on the $\mathcal{L}$ and $\mathcal{K}$ operators, if the environment is homogeneous isotropic lossless. Taking the GFHF (3) as an example, it can be equivalently rewritten as follows:

$$\left.\begin{array}{l}\text{ext}V^{mat}_{sim}:-\vec{E}^{sca}\\ \text{int}V^{mat}_{sim}:\vec{E}^{inc}\end{array}\right\}=-j\omega\mu_0\mathcal{L}_0(\vec{J}^{ES})-\mathcal{K}_0(\vec{M}^{ES}) \quad (3.1')$$

$$\left.\begin{array}{l}\text{ext}V^{mat}_{sim}:-\vec{H}^{sca}\\ \text{int}V^{mat}_{sim}:\vec{H}^{inc}\end{array}\right\}=-j\omega\varepsilon_0\mathcal{L}_0(\vec{M}^{ES})+\mathcal{K}_0(\vec{J}^{ES}) \quad (3.2')$$

if the environment is vacuum. In (3.1') and (3.2'), the operators $\mathcal{L}_0$ and $\mathcal{K}_0$ are defined as follows: [45]

$$\mathcal{L}_0(\vec{X}) \triangleq \left(1+\frac{1}{k_0^2}\nabla\nabla\cdot\right)\int_\Pi G_0(\vec{r},\vec{r}')\vec{X}(\vec{r}')d\Pi' \quad (16.1)$$

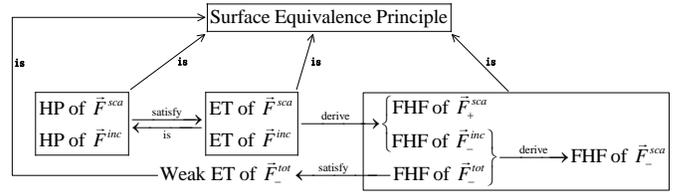

Fig. 6. The relationships among HP, ET, FHF, and surface equivalence principle.

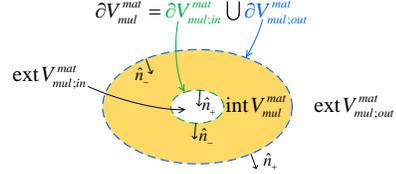

Fig. 7. Various domains related to a 2-connected material body.

$$\mathcal{K}_0(\vec{X}) \triangleq \nabla\times\int_\Pi G_0(\vec{r},\vec{r}')\vec{X}(\vec{r}')d\Pi' \quad (16.2)$$

where $k_0 = \omega\sqrt{\mu_0\varepsilon_0}$, and $G_0(\vec{r},\vec{r}') = e^{-jk_0|\vec{r}-\vec{r}'|}/4\pi|\vec{r}-\vec{r}'|$.

**Application on constructing CM**

Recently, it is clarified in [46] that: i) the physical essence of CM theory [35], [47]-[49] is to orthogonalize the power done by incident field on scattering current; ii) the arguments of the power operator in CM theory must be independent of each other; and iii) the equivalent currents $\vec{J}^{ES}$ and $\vec{M}^{ES}$ on $\partial V^{mat}_{sim}$ depend on each other.

Formulation (5) expresses the $\vec{F}^{tot}$ in terms of $\{\vec{J}^{ES}, \vec{M}^{ES}\}$, and then the scattering currents $\{\vec{J}^{SV}, \vec{M}^{SV}\}$ can be expressed in terms of $\{\vec{J}^{ES}, \vec{M}^{ES}\}$, because $\vec{J}^{SV} = j\omega\Delta\vec{\varepsilon}_c \cdot \vec{E}^{tot}_-$ and $\vec{M}^{SV} = j\omega\Delta\vec{\mu} \cdot \vec{H}^{tot}_-$ [1], [2], [45]. Based on this observation and formulation (3), the power done by incident field on scattering current can be efficiently expressed in terms of $\{\vec{J}^{ES}, \vec{M}^{ES}\}$. Because of the continuity of the tangential $\vec{F}^{sca}$ on $\partial V^{mat}_{sim}$, the $\vec{J}^{ES}$ and $\vec{M}^{ES}$ can be expressed in terms of each other, based on formulations (3) and (6) and the method given in [46]. Because of these above, it can be concluded that the GFHF is valuable for constructing the CM of inhomogeneous anisotropic lossy material body.

### III. A Single Multiply Connected Inhomogeneous Anisotropic Lossy Material Body

In this section, the results obtained in above Sec. II are generalized to the EM system $V^{mat}_{sys}$ which is a multiply connected inhomogeneous anisotropic lossy material body $V^{mat}_{mul}$, and the $V^{mat}_{mul}$ is restricted to being 2-connected as shown in Fig. 7. The arbitrary $l$-connected case ($l > 2$) can be similarly discussed, and the corresponding formulations are identical to the 2-connected case in form. Then, $V^{mat}_{sim} = \varnothing$, and $V^{mat}_{sys} = V^{mat}_{mul}$, and $\vec{F}^{sca} = \vec{F}^{sca}_{mul}$, and $\vec{F}^{tot} = \vec{F}^{inc} + \vec{F}^{sca} = \vec{F}^{inc} + \vec{F}^{sca}_{mul}$, in this section.

The whole boundary $\partial V^{mat}_{mul}$ and whole $\text{ext}V^{mat}_{mul}$ can be decomposed as follows: [38]

$$\partial V^{mat}_{mul} = \partial V^{mat}_{mul;in} \cup \partial V^{mat}_{mul;out} \quad (17.1)$$

$$\text{ext} V_{mul}^{mat} = \text{ext} V_{mul;in}^{mat} \cup \text{ext} V_{mul;out}^{mat} \quad (17.2)$$

where $\partial V_{mul;in}^{mat}$, $\partial V_{mul;out}^{mat}$, $\text{ext} V_{mul;in}^{mat}$, and $\text{ext} V_{mul;out}^{mat}$ are shown in Fig. 7. In this section, it is restricted that the $\{\vec{J}^{inc}, \vec{M}^{inc}\}$ distribute on $\text{ext} V_{mul;out}^{mat}$, and the case that $\{\vec{J}^{inc}, \vec{M}^{inc}\}$ distribute on $\text{ext} V_{mul;in}^{mat}$ can be similarly discussed, and the final formulations of two different cases are identical to each other in form. The composite case corresponding to that $\{\vec{J}^{inc}, \vec{M}^{inc}\}$ simultaneously distribute on $\text{ext} V_{mul;out}^{mat}$ and $\text{ext} V_{mul;in}^{mat}$ can be viewed as the superposition of two simple cases, based on superposition principle [50], and the final formulations of composite case are identical to the simple cases in form.

*A. Generalized Huygens' principle and extinction theorem*

Similarly to deriving (1') and (2), for multiply connected material system the following formulations can be derived:

$$\left.\begin{array}{rcl}\text{ext} V_{mul;out}^{mat} &:& 0 \\ \text{int} V_{mul}^{mat} &:& 0 \\ \text{ext} V_{mul;in}^{mat} &:& -\vec{F}^{inc}\end{array}\right\} = \left[\ddot{\vec{G}}_{env}^{JF} * (\hat{n}_{-} \times \vec{H}^{inc}) + \ddot{\vec{G}}_{env}^{MF} * (\vec{E}^{inc} \times \hat{n}_{-})\right]_{\partial V_{mul;in}^{mat}}$$
(18.1)

$$\left.\begin{array}{rcl}\text{ext} V_{mul;out}^{mat} &:& 0 \\ \text{int} V_{mul}^{mat} &:& 0 \\ \text{ext} V_{mul;in}^{mat} &:& \vec{F}^{sca}\end{array}\right\} = \left[\ddot{\vec{G}}_{env}^{JF} * (\hat{n}_{+} \times \vec{H}^{sca}) + \ddot{\vec{G}}_{env}^{MF} * (\vec{E}^{sca} \times \hat{n}_{+})\right]_{\partial V_{mul;in}^{mat}}$$
(19.1)

if the Huygens' surface is selected as $\partial V_{mul;in}^{mat}$; the following formulations can be derived:

$$\left.\begin{array}{rcl}\text{ext} V_{mul;out}^{mat} &:& 0 \\ \text{int} V_{mul}^{mat} &:& \vec{F}^{inc} \\ \text{ext} V_{mul;in}^{mat} &:& \vec{F}^{inc}\end{array}\right\} = \left[\ddot{\vec{G}}_{env}^{JF} * (\hat{n}_{-} \times \vec{H}^{inc}) + \ddot{\vec{G}}_{env}^{MF} * (\vec{E}^{inc} \times \hat{n}_{-})\right]_{\partial V_{mul;out}^{mat}}$$
(18.2)

$$\left.\begin{array}{rcl}\text{ext} V_{mul;out}^{mat} &:& \vec{F}^{sca} \\ \text{int} V_{mul}^{mat} &:& 0 \\ \text{ext} V_{mul;in}^{mat} &:& 0\end{array}\right\} = \left[\ddot{\vec{G}}_{env}^{JF} * (\hat{n}_{+} \times \vec{H}^{sca}) + \ddot{\vec{G}}_{env}^{MF} * (\vec{E}^{sca} \times \hat{n}_{+})\right]_{\partial V_{mul;out}^{mat}}$$
(19.2)

if the Huygens' surface is selected as $\partial V_{mul;out}^{mat}$. The $\hat{n}_{+}$ and $\hat{n}_{-}$ in (18) and (19) are shown in Fig. 7. The summation of (18.1) and (18.2) and the summation of (19.1) and (19.2) are that

$$\left.\begin{array}{rcl}\text{ext} V_{mul}^{mat} &:& 0 \\ \text{int} V_{mul}^{mat} &:& \vec{F}^{inc}\end{array}\right\} = \left[\ddot{\vec{G}}_{env}^{JF} * (\hat{n}_{-} \times \vec{H}^{inc}) + \ddot{\vec{G}}_{env}^{MF} * (\vec{E}^{inc} \times \hat{n}_{-})\right]_{\partial V_{mul}^{mat}} \quad (18')$$

$$\left.\begin{array}{rcl}\text{ext} V_{mul}^{mat} &:& \vec{F}^{sca} \\ \text{int} V_{mul}^{mat} &:& 0\end{array}\right\} = \left[\ddot{\vec{G}}_{env}^{JF} * (\hat{n}_{+} \times \vec{H}^{sca}) + \ddot{\vec{G}}_{env}^{MF} * (\vec{E}^{sca} \times \hat{n}_{+})\right]_{\partial V_{mul}^{mat}} \quad (19')$$

in which (17) has been utilized to simplify the symbolic expressions of integral domain, interesting domain, and excluded domain.

In Sec. II, it has been clarified that the physical essence of ETs (1') and (2) is the HP corresponding to a simply connected inhomogeneous anisotropic lossy material body. As the counterparts of (1') and (2), the *generalized extinction theorems (GETs)* (18') and (19') can be viewed as the mathematical formulation of the *generalized Huygens' principle (GHP)* of multiply connected case. The adjective "generalized" is due to that: the Huygens' surface used in traditional HP and ET is a single closed surface; however, the "Huygens' surface" in these generalized versions is constructed by multiple closed surfaces, as shown in Fig. 7. In fact, the "Huygens' surface" will be further generalized to so-called "Huygens' boundary" that includes some lines and open surfaces besides closed surfaces, in our future works. In addition, it should be emphasized that the GHP (18') and the GHP (19') satisfy Hadamard's syllogism and all the conditions listed in Sec. II-C.

*B. Generalized Franz-Harrington formulation and weak extinction theorem*

Similarly to deriving (3) from (1') and (2), the following formulation (20) can be derived from the above (18') and (19'):

$$\left.\begin{array}{rcl}\text{ext} V_{mul}^{mat} &:& -\vec{F}^{sca} \\ \text{int} V_{mul}^{mat} &:& \vec{F}^{inc}\end{array}\right\} = \left[\ddot{\vec{G}}_{env}^{JF} * \vec{J}^{ES} + \ddot{\vec{G}}_{env}^{MF} * \vec{M}^{ES}\right]_{\partial V_{mul}^{mat}}. \quad (20)$$

In (20), the $\vec{J}^{ES}$ and $\vec{M}^{ES}$ are defined as (4), except that the material boundary should be replaced by $\partial V_{mul}^{mat}$.

Similarly to generalizing (3) to (20), (5) can be generalized to the following (21):

$$\text{int} V_{mul}^{mat} \quad : \quad \vec{F}^{tot} = \left[\ddot{\vec{G}}_{mul}^{JF} * \vec{J}^{ES} + \ddot{\vec{G}}_{mul}^{MF} * \vec{M}^{ES}\right]_{\partial V_{mul}^{mat}} \quad (21)$$

where $\ddot{\vec{G}}_{mul}^{JF}(\vec{r},\vec{r}')$ and $\ddot{\vec{G}}_{mul}^{MF}(\vec{r},\vec{r}')$ are the Green's functions corresponding to material body $V_{mul}^{mat}$, and then

$$\text{int} V_{mul}^{mat} \quad : \quad \vec{F}^{sca} = \left[\Delta\ddot{\vec{G}}_{mul}^{JF} * \vec{J}^{ES} + \Delta\ddot{\vec{G}}_{mul}^{MF} * \vec{M}^{ES}\right]_{\partial V_{mul}^{mat}} \quad (22)$$

based on (20) and (21), where $\Delta\ddot{\vec{G}}_{mul}^{JF}(\vec{r},\vec{r}')$ and $\Delta\ddot{\vec{G}}_{mul}^{MF}(\vec{r},\vec{r}')$ are defined similarly to (7).

## IV. MULTIPLE CONNECTED INHOMOGENEOUS ANISOTROPIC LOSSY MATERIAL BODIES

The above Secs. II and III only focus on the EM system $V_{sys}^{mat}$ which is constructed by a single material body. In this section, the $V_{sys}^{mat}$ constructed by two bodies are considered, and the $V_{sys}^{mat}$ constructed by arbitrary $l$ ($l>2$) bodies can be similarly discussed. Some typical examples of two-body system $V_{sys}^{mat}$ are shown in Fig. 8, and they include a simply connected body $V_{sim}^{mat}$ and a multiply connected body $V_{mul}^{mat}$, i.e., $V_{sys}^{mat} = V_{sim}^{mat} \cup V_{mul}^{mat}$. The case that $V_{sys}^{mat}$ is constructed by two simply connected bodies and the case that $V_{sys}^{mat}$ is constructed by two multiply connected bodies can be similarly discussed, and their results are identical to the results of case $V_{sys}^{mat} = V_{sim}^{mat} \cup V_{mul}^{mat}$ in form.

In the following subsections, the current decomposition method (CDM) is developed at first, and then the results obtained in Secs. II and III are generalized to two-body system.

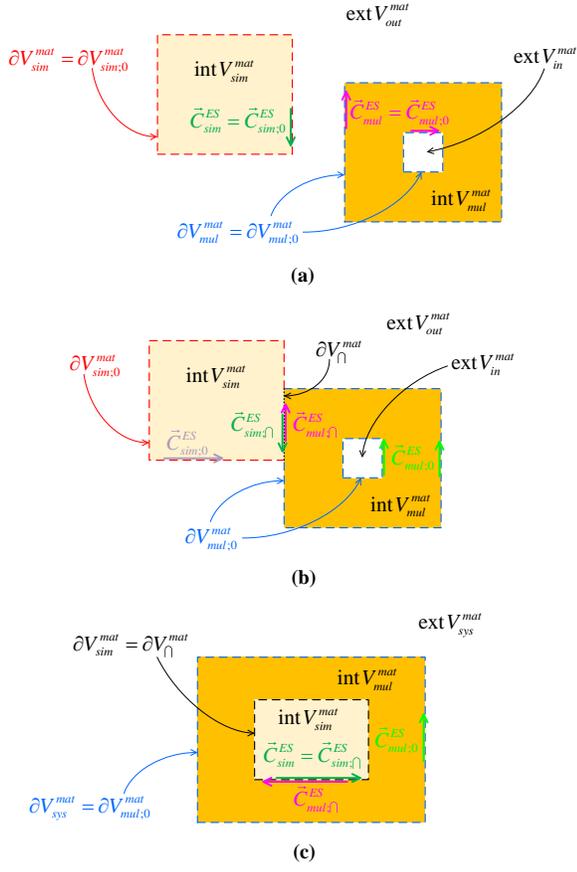

Fig. 8. (a) Various domains related to a non-connected two-body system; (b) various domains related to a two-body system, in which a body contacts with but doesn't submerge into another body; (c) various domains related to a two-body system, in which a body is submerged into another body.

### A. Current decomposition method

The boundaries of $V_{sim}^{mat}$ and $V_{mul}^{mat}$ can be decomposed as

$$\partial V_{sim/mul}^{mat} = \partial V_{sim/mul;0}^{mat} \cup \partial V_{\cap}^{mat} \quad (23)$$

where

$$\partial V_{sim/mul;0}^{mat} \triangleq \text{cl}\left(\partial V_{sim/mul}^{mat} \setminus \partial V_{mul/sim}^{mat}\right) \quad (24.1)$$

$$\partial V_{\cap}^{mat} \triangleq \partial V_{sim}^{mat} \setminus \partial V_{sim;0}^{mat}$$
$$= \partial V_{mul}^{mat} \setminus \partial V_{mul;0}^{mat} \quad (24.2)$$

as shown in Fig. 8. Obviously, the $\partial V_{sim/mul;0}^{mat}$ and $\partial V_{\cap}^{mat}$ are disjoint, i.e.,

$$\partial V_{sim/mul;0}^{mat} \cap \partial V_{\cap}^{mat} = \varnothing. \quad (25)$$

Based on (23) and (25), the equivalent surface currents on $\partial V_{sim}^{mat}$ and $\partial V_{mul}^{mat}$ can be decomposed as follows:

$$\vec{C}_{sim/mul}^{ES}(\vec{r}) = \vec{C}_{sim/mul;0}^{ES}(\vec{r}) + \vec{C}_{sim/mul;\cap}^{ES}(\vec{r}) \quad , \quad (\vec{r} \in \partial V_{sim/mul}^{mat}) \quad (26)$$

where $C = J, M$, and

$$\vec{C}_{sim/mul;0}^{ES}(\vec{r}) \triangleq \begin{cases} \vec{C}_{sim/mul}^{ES}(\vec{r}) &, \quad (\vec{r} \in \partial V_{sim/mul;0}^{mat}) \\ 0 &, \quad (\vec{r} \in \partial V_{\cap}^{mat}) \end{cases} \quad (27.1)$$

$$\vec{C}_{sim/mul;\cap}^{ES}(\vec{r}) \triangleq \begin{cases} 0 &, \quad (\vec{r} \in \partial V_{sim/mul;0}^{mat}) \\ \vec{C}_{sim/mul}^{ES}(\vec{r}) &, \quad (\vec{r} \in \partial V_{\cap}^{mat}) \end{cases}. \quad (27.2)$$

In (27), the $\vec{C}_{sim/mul}^{ES}$ is defined similarly to (4), except that the material boundary in (4) should be replaced by $\partial V_{sim/mul}^{mat}$. Because the polarization electric current and magnetization magnetic current models are utilized to depict the polarization and magnetization phenomena in this paper, there doesn't exist any scattering [1], [45] and incident surface current on $\partial V_{sim/mul}^{mat}$, and then the tangential components of total field $\vec{F}^{tot}$ are continuous on $\partial V_{\cap}^{mat}$. Hence, the following relationship exists for any $C = J, M$:

$$\vec{C}_{sim;\cap}^{ES}(\vec{r}) = -\vec{C}_{mul;\cap}^{ES}(\vec{r}) \quad , \quad (\vec{r} \in \partial V_{\cap}^{mat}). \quad (28)$$

### B. Generalized Huygens' principle, extinction theorem, and Franz-Harrington formulation: General case

In this subsection, the results obtained in Secs. II and III are generalized to a general two-body inhomogeneous anisotropic lossy system in inhomogeneous anisotropic lossy environment.

**Generalized Huygens' principle and extinction theorem**
Based on (1') and (2), we have that

$$\left. \begin{array}{l} \text{ext}\, V_{sys}^{mat} \; : \; 0 \\ \text{int}\, V_{sim}^{mat} \; : \; \vec{F}^{inc} \\ \text{int}\, V_{mul}^{mat} \; : \; 0 \end{array} \right\} = \left[ \ddot{G}_{env}^{JF} * \left(\hat{n}_{sim;-} \times \vec{H}^{inc}\right) + \ddot{G}_{env}^{MF} * \left(\vec{E}^{inc} \times \hat{n}_{sim;-}\right) \right]_{\partial V_{sim;0}^{mat} \cup \partial V_{\cap}^{mat}} \quad (29.1)$$

$$\left. \begin{array}{l} \text{ext}\, V_{sys}^{mat} \; : \; \vec{F}_{sim}^{sca} \\ \text{int}\, V_{sim}^{mat} \; : \; 0 \\ \text{int}\, V_{mul}^{mat} \; : \; \vec{F}_{sim}^{sca} \end{array} \right\} = \left[ \ddot{G}_{env}^{JF} * \left(\hat{n}_{sim;+} \times \vec{H}_{sim}^{sca}\right) + \ddot{G}_{env}^{MF} * \left(\vec{E}_{sim}^{sca} \times \hat{n}_{sim;+}\right) \right]_{\partial V_{sim;0}^{mat} \cup \partial V_{\cap}^{mat}} \quad (30.1)$$

where $\hat{n}_{sim;-}$ and $\hat{n}_{sim;+}$ are the normal vectors of $\partial V_{sim}^{mat}$, and respectively point to the interior and exterior of $V_{sim}^{mat}$. Based on (18') and (19'), we have that

$$\left. \begin{array}{l} \text{ext}\, V_{sys}^{mat} \; : \; 0 \\ \text{int}\, V_{sim}^{mat} \; : \; 0 \\ \text{int}\, V_{mul}^{mat} \; : \; \vec{F}^{inc} \end{array} \right\} = \left[ \ddot{G}_{env}^{JF} * \left(\hat{n}_{mul;-} \times \vec{H}^{inc}\right) + \ddot{G}_{env}^{MF} * \left(\vec{E}^{inc} \times \hat{n}_{mul;-}\right) \right]_{\partial V_{mul;0}^{mat} \cup \partial V_{\cap}^{mat}} \quad (29.2)$$

$$\left. \begin{array}{l} \text{ext}\, V_{sys}^{mat} \; : \; \vec{F}_{mul}^{sca} \\ \text{int}\, V_{sim}^{mat} \; : \; \vec{F}_{mul}^{sca} \\ \text{int}\, V_{mul}^{mat} \; : \; 0 \end{array} \right\} = \left[ \ddot{G}_{env}^{JF} * \left(\hat{n}_{mul;+} \times \vec{H}_{mul}^{sca}\right) + \ddot{G}_{env}^{MF} * \left(\vec{E}_{mul}^{sca} \times \hat{n}_{mul;+}\right) \right]_{\partial V_{mul;0}^{mat} \cup \partial V_{\cap}^{mat}} \quad (30.2)$$

where $\hat{n}_{mul;-}$ and $\hat{n}_{mul;+}$ are the normal vectors of $\partial V_{mul}^{mat}$, and respectively point to the interior and exterior of $V_{mul}^{mat}$.

In addition, we also have that

$$\left.\begin{array}{ll}\text{ext}V_{sys}^{mat} & : \quad 0 \\ \text{int}V_{sim}^{mat} & : \quad -\vec{F}_{mul}^{sca} \\ \text{int}V_{mul}^{mat} & : \quad 0\end{array}\right\} = \left[\ddot{G}_{env}^{JF} * \left(\hat{n}_{sim;+} \times \vec{H}_{mul}^{sca}\right) + \ddot{G}_{env}^{MF} * \left(\vec{E}_{mul}^{sca} \times \hat{n}_{sim;+}\right)\right]_{\partial V_{sim;0}^{mat} \cup \partial V_{\cap}^{mat}}$$

(30.3)

$$\left.\begin{array}{ll}\text{ext}V_{sys}^{mat} & : \quad 0 \\ \text{int}V_{sim}^{mat} & : \quad 0 \\ \text{int}V_{mul}^{mat} & : \quad -\vec{F}_{sim}^{sca}\end{array}\right\} = \left[\ddot{G}_{env}^{JF} * \left(\hat{n}_{mul;+} \times \vec{H}_{sim}^{sca}\right) + \ddot{G}_{env}^{MF} * \left(\vec{E}_{sim}^{sca} \times \hat{n}_{mul;+}\right)\right]_{\partial V_{mul;0}^{mat} \cup \partial V_{\cap}^{mat}}$$

(30.4)

based on the method similarly to deriving (1') and (18').

The summation of (29.1) and (29.2) gives that

$$\left.\begin{array}{ll}\text{ext}V_{sys}^{mat} & : \quad 0 \\ \text{int}V_{sim}^{mat} & : \quad \vec{F}^{inc} \\ \text{int}V_{mul}^{mat} & : \quad \vec{F}^{inc}\end{array}\right\} = \left[\ddot{G}_{env}^{JF} * \left(\hat{n}_{sim;-} \times \vec{H}^{inc}\right) + \ddot{G}_{env}^{MF} * \left(\vec{E}^{inc} \times \hat{n}_{sim;-}\right)\right]_{\partial V_{sim;0}^{mat} \cup \partial V_{\cap}^{mat}} \\ + \left[\ddot{G}_{env}^{JF} * \left(\hat{n}_{mul;-} \times \vec{H}^{inc}\right) + \ddot{G}_{env}^{MF} * \left(\vec{E}^{inc} \times \hat{n}_{mul;-}\right)\right]_{\partial V_{mul;0}^{mat} \cup \partial V_{\cap}^{mat}}$$

(29')

and the summation of (30.1)-(30.4) gives that

$$\left.\begin{array}{ll}\text{ext}V_{sys}^{mat} & : \quad \vec{F}^{sca} \\ \text{int}V_{sim}^{mat} & : \quad 0 \\ \text{int}V_{mul}^{mat} & : \quad 0\end{array}\right\} = \left[\ddot{G}_{env}^{JF} * \left(\hat{n}_{sim;+} \times \vec{H}^{sca}\right) + \ddot{G}_{env}^{MF} * \left(\vec{E}^{sca} \times \hat{n}_{sim;+}\right)\right]_{\partial V_{sim;0}^{mat} \cup \partial V_{\cap}^{mat}} \\ + \left[\ddot{G}_{env}^{JF} * \left(\hat{n}_{mul;+} \times \vec{H}^{sca}\right) + \ddot{G}_{env}^{MF} * \left(\vec{E}^{sca} \times \hat{n}_{mul;+}\right)\right]_{\partial V_{mul;0}^{mat} \cup \partial V_{\cap}^{mat}}$$

(30')

The above (29') and (30') are called as the *topological additivity* of GHP and GET, i.e., the GHP/GET of whole EM system equals to the summation of the GHP/GET corresponding to all sub-systems as formulated in following (31) and (32), and this property is consistent with the principle of superposition.

Scattering field GHP/GET of whole material system
$= \sum_{\xi}$ Scattering field GHP/GET of material body $V_{\xi}^{mat}$ (31)

Incident field GHP/GET of whole material system
$= \sum_{\xi}$ Incident field GHP/GET of material body $V_{\xi}^{mat}$. (32)

**Generalized Franz-Harrington formulation**

Similarly to deriving (3) from (1') and (2) and deriving (20) from (18') and (19'), the following (33) can be derived from (29') and (30'):

$$\left.\begin{array}{ll}\text{ext}V_{sys}^{mat} & : \quad -\vec{F}^{sca} \\ \text{int}V_{sim}^{mat} & : \quad \vec{F}^{inc} \\ \text{int}V_{mul}^{mat} & : \quad \vec{F}^{inc}\end{array}\right\} = \left[\ddot{G}_{env}^{JF} * \left(\vec{J}_{sim;0}^{ES} + \vec{J}_{sim;\cap}^{ES}\right) + \ddot{G}_{env}^{MF} * \left(\vec{M}_{sim;0}^{ES} + \vec{M}_{sim;\cap}^{ES}\right)\right]_{\partial V_{sim;0}^{mat} \cup \partial V_{\cap}^{mat}} \\ + \left[\ddot{G}_{env}^{JF} * \left(\vec{J}_{mul;0}^{ES} + \vec{J}_{mul;\cap}^{ES}\right) + \ddot{G}_{env}^{MF} * \left(\vec{M}_{mul;0}^{ES} + \vec{M}_{mul;\cap}^{ES}\right)\right]_{\partial V_{mul;0}^{mat} \cup \partial V_{\cap}^{mat}}$$

(33)

where (23) and (26) have been utilized. Obviously, the GFHF of internal incident field and external scattering field satisfy topological additivity (34.1) and (34.2) just like GHP and GET, because they are essentially the summation of incident field GHP and scattering field GHP as pointed out in Sec. II-C.

Internal incident field GFHF of whole material system
$= \sum_{\xi}$ Internal incident field GFHF of material body $V_{\xi}^{mat}$ (34.1)

External scattering field GFHF of whole material system
$= \sum_{\xi}$ External scattering field GFHF of material body $V_{\xi}^{mat}$. (34.2)

However, the GFHF of internal total field and internal scattering field don't satisfy topological additivity, because they don't satisfy GET. To resolve this problem, the following *piecewise Green's functions* are proposed:

$$\tilde{\ddot{G}}_{sim/mul}^{JF}(\vec{r},\vec{r}') \triangleq \begin{cases} \ddot{G}_{sim/mul}^{JF}(\vec{r},\vec{r}') & , \left(\vec{r} \in \text{cl}V_{sim/mul}^{mat}, \vec{r}' \in \text{cl}V_{sim/mul}^{mat}\right) \\ 0 & , \left(\vec{r} \in \text{ext}V_{sim/mul}^{mat}, \vec{r}' \in \text{cl}V_{sim/mul}^{mat}\right)\end{cases}$$ (35.1)

$$\tilde{\ddot{G}}_{sim/mul}^{MF}(\vec{r},\vec{r}') \triangleq \begin{cases} \ddot{G}_{sim/mul}^{MF}(\vec{r},\vec{r}') & , \left(\vec{r} \in \text{cl}V_{sim/mul}^{mat}, \vec{r}' \in \text{cl}V_{sim/mul}^{mat}\right) \\ 0 & , \left(\vec{r} \in \text{ext}V_{sim/mul}^{mat}, \vec{r}' \in \text{cl}V_{sim/mul}^{mat}\right)\end{cases}$$ (35.2)

Based on (35), (5) and (21) can be rewritten as follows:

$$\left.\begin{array}{ll}\text{ext}V_{sim}^{mat} & : \quad 0 \\ \text{int}V_{sim}^{mat} & : \quad \vec{F}^{tot}\end{array}\right\} = \left[\tilde{\ddot{G}}_{sim}^{JF} * \vec{J}^{ES} + \tilde{\ddot{G}}_{sim}^{MF} * \vec{M}^{ES}\right]_{\partial V_{sim}^{mat}}$$ (5')

$$\left.\begin{array}{ll}\text{ext}V_{mul}^{mat} & : \quad 0 \\ \text{int}V_{mul}^{mat} & : \quad \vec{F}^{tot}\end{array}\right\} = \left[\tilde{\ddot{G}}_{mul}^{JF} * \vec{J}^{ES} + \tilde{\ddot{G}}_{mul}^{MF} * \vec{M}^{ES}\right]_{\partial V_{mul}^{mat}}$$ (21')

and then (5') and (21') can be generalized to the following (36):

$$\left.\begin{array}{ll}\text{ext}V_{sys}^{mat} & : \quad 0 \\ \text{int}V_{sim}^{mat} & : \quad \vec{F}^{tot} \\ \text{int}V_{mul}^{mat} & : \quad \vec{F}^{tot}\end{array}\right\} = \left[\tilde{\ddot{G}}_{sim}^{JF} * \left(\vec{J}_{sim;0}^{ES} + \vec{J}_{sim;\cap}^{ES}\right) + \tilde{\ddot{G}}_{sim}^{MF} * \left(\vec{M}_{sim;0}^{ES} + \vec{M}_{sim;\cap}^{ES}\right)\right]_{\partial V_{sim;0}^{mat} \cup \partial V_{\cap}^{mat}} \\ + \left[\tilde{\ddot{G}}_{mul}^{JF} * \left(\vec{J}_{mul;0}^{ES} + \vec{J}_{mul;\cap}^{ES}\right) + \tilde{\ddot{G}}_{mul}^{MF} * \left(\vec{M}_{mul;0}^{ES} + \vec{M}_{mul;\cap}^{ES}\right)\right]_{\partial V_{mul;0}^{mat} \cup \partial V_{\cap}^{mat}}$$

(36)

In fact, (5'), (21'), and (36) can be called as *artificial extinction theorems* to be distinguished from extinction theorem and weak extinction theorem.

Similarly, if the following *delta piecewise Green's functions* are proposed:

$$\Delta\tilde{\ddot{G}}_{sim/mul}^{JF}(\vec{r},\vec{r}') \triangleq \tilde{\ddot{G}}_{sim/mul}^{JF}(\vec{r},\vec{r}') - \ddot{G}_{env}^{JF}(\vec{r},\vec{r}') \\ = \begin{cases} \ddot{G}_{sim/mul}^{JF}(\vec{r},\vec{r}') - \ddot{G}_{env}^{JF}(\vec{r},\vec{r}') & , \left(\vec{r} \in \text{cl}V_{sim/mul}^{mat}, \vec{r}' \in \text{cl}V_{sim/mul}^{mat}\right) \\ -\ddot{G}_{env}^{JF}(\vec{r},\vec{r}') & , \left(\vec{r} \in \text{ext}V_{sim/mul}^{mat}, \vec{r}' \in \text{cl}V_{sim/mul}^{mat}\right)\end{cases}$$

(7.1')

$$\Delta\tilde{\ddot{G}}_{sim/mul}^{MF}(\vec{r},\vec{r}') \triangleq \tilde{\ddot{G}}_{sim/mul}^{MF}(\vec{r},\vec{r}') - \ddot{G}_{env}^{MF}(\vec{r},\vec{r}') \\ = \begin{cases} \ddot{G}_{sim/mul}^{MF}(\vec{r},\vec{r}') - \ddot{G}_{env}^{MF}(\vec{r},\vec{r}') & , \left(\vec{r} \in \text{cl}V_{sim/mul}^{mat}, \vec{r}' \in \text{cl}V_{sim/mul}^{mat}\right) \\ -\ddot{G}_{env}^{MF}(\vec{r},\vec{r}') & , \left(\vec{r} \in \text{ext}V_{sim/mul}^{mat}, \vec{r}' \in \text{cl}V_{sim/mul}^{mat}\right)\end{cases}$$

(7.2')

(6) and (22) can be rewritten as follows:

$$\left.\begin{array}{ll}\text{ext}V_{sim}^{mat} & : \quad \vec{F}_{sim}^{sca} \\ \text{int}V_{sim}^{mat} & : \quad \vec{F}_{sim}^{sca}\end{array}\right\} = \left[\Delta\tilde{\ddot{G}}_{sim}^{JF} * \vec{J}^{ES} + \Delta\tilde{\ddot{G}}_{sim}^{MF} * \vec{M}^{ES}\right]_{\partial V_{sim}^{mat}}$$ (6')

$$\left.\begin{array}{ll}\text{ext}V_{mul}^{mat} & : \quad \vec{F}_{mul}^{sca} \\ \text{int}V_{mul}^{mat} & : \quad \vec{F}_{mul}^{sca}\end{array}\right\} = \left[\Delta\tilde{\ddot{G}}_{mul}^{JF} * \vec{J}^{ES} + \Delta\tilde{\ddot{G}}_{mul}^{MF} * \vec{M}^{ES}\right]_{\partial V_{mul}^{mat}}$$ (22')

and then (6') and (22') can be generalized to the following (37):

$$\left.\begin{array}{rl} \text{ext}V_{sys}^{mat} & : \vec{F}^{sca} \\ \text{int}V_{sim}^{mat} & : \vec{F}^{sca} \\ \text{int}V_{mul}^{mat} & : \vec{F}^{sca} \end{array}\right\} = \left[\Delta\tilde{\vec{G}}_{sim}^{JF} * \left(\vec{J}_{sim;0}^{ES} + \vec{J}_{sim;\cap}^{ES}\right) + \Delta\tilde{\vec{G}}_{sim}^{MF} * \left(\vec{M}_{sim;0}^{ES} + \vec{M}_{sim;\cap}^{ES}\right)\right]_{\partial V_{sim;0}^{mat} \cup \partial V_{\cap}^{mat}} + \left[\Delta\tilde{\vec{G}}_{mul}^{JF} * \left(\vec{J}_{mul;0}^{ES} + \vec{J}_{mul;\cap}^{ES}\right) + \Delta\tilde{\vec{G}}_{mul}^{MF} * \left(\vec{M}_{mul;0}^{ES} + \vec{M}_{mul;\cap}^{ES}\right)\right]_{\partial V_{mul;0}^{mat} \cup \partial V_{\cap}^{mat}} \quad (37)$$

by summing (6') and (22').

Obviously, the piecewise-Green-function-based GFHF of internal total field and internal scattering field satisfy the following topological additivity (38) and (39) just like the GFHF of internal incident field and external scattering field.

$$\begin{array}{l}\text{Internal total field GFHF of whole material system} \\ = \sum_{\xi} \text{Internal total field GFHF of material body } V_{\xi}^{mat}\end{array} \quad (38)$$

$$\begin{array}{l}\text{Internal scattering field GFHF of whole material system} \\ = \sum_{\xi} \text{Internal scattering field GFHF of material body } V_{\xi}^{mat}\end{array} \quad (39)$$

In fact, (34), (38), and (39) can be uniformly written as

$$\begin{array}{l}\text{The GFHF of whole material system} \\ = \sum_{\xi} \text{The GFHF of material body } V_{\xi}^{mat}\end{array} \quad (40)$$

(31), (32), and (40) are called as the topological additivity of GHP, GET, and GFHF.

*C. Generalized Huygens' principle, extinction theorem, and Franz-Harrington formulation: Special cases shown in Fig. 8*

In this subsection, the results for a general two-body material system are specialized to some special cases.

**Case I: Two bodies don't contact with each other**

In this case, the following relationships exist:

$$\partial V_{sys}^{mat} = \partial V_{sim}^{mat} \cup \partial V_{mul}^{mat} \quad (41.1)$$
$$\text{ext}V_{sys}^{mat} = \text{ext}V_{in}^{mat} \cup \text{ext}V_{out}^{mat} \quad (41.2)$$
$$\text{int}V_{sys}^{mat} = \text{int}V_{sim}^{mat} \cup \text{int}V_{mul}^{mat} \quad (41.3)$$

where the $\partial V_{sim}^{mat}$, $\partial V_{mul}^{mat}$, $\text{ext}V_{in}^{mat}$, $\text{ext}V_{out}^{mat}$, $\text{int}V_{sim}^{mat}$, and $\text{int}V_{mul}^{mat}$ are shown in Fig. 8 (a).

Then, the incident field GHP (29') and scattering field GHP (30') are specialized to

$$\left.\begin{array}{rl} \text{ext}V_{sys}^{mat} & : 0 \\ \text{int}V_{sys}^{mat} & : \vec{F}^{inc} \end{array}\right\} = \left[\tilde{\vec{G}}_{env}^{JF} * \left(\hat{n}_{-} \times \vec{H}^{inc}\right) + \tilde{\vec{G}}_{env}^{MF} * \left(\vec{E}^{inc} \times \hat{n}_{-}\right)\right]_{\partial V_{sys}^{mat}} \quad (42)$$

$$\left.\begin{array}{rl} \text{ext}V_{sys}^{mat} & : \vec{F}^{sca} \\ \text{int}V_{sys}^{mat} & : 0 \end{array}\right\} = \left[\tilde{\vec{G}}_{env}^{JF} * \left(\hat{n}_{+} \times \vec{H}^{sca}\right) + \tilde{\vec{G}}_{env}^{MF} * \left(\vec{E}^{sca} \times \hat{n}_{+}\right)\right]_{\partial V_{sys}^{mat}} \quad (43)$$

and the GFHF (33) is specialized to

$$\left.\begin{array}{rl} \text{ext}V_{sys}^{mat} & : -\vec{F}^{sca} \\ \text{int}V_{sys}^{mat} & : \vec{F}^{inc} \end{array}\right\} = \left[\tilde{\vec{G}}_{env}^{JF} * \vec{J}^{ES} + \tilde{\vec{G}}_{env}^{MF} * \vec{M}^{ES}\right]_{\partial V_{sys}^{mat}} \quad (44)$$

and the GFHFs (36) and (37) are specialized to

$$\left.\begin{array}{rl} \text{ext}V_{sys}^{mat} & : 0 \\ \text{int}V_{sys}^{mat} & : \vec{F}^{tot} \end{array}\right\} = \left[\tilde{\vec{G}}_{sys}^{JF} * \vec{J}^{ES} + \tilde{\vec{G}}_{sys}^{MF} * \vec{M}^{ES}\right]_{\partial V_{sys}^{mat}} \quad (45)$$

$$\left.\begin{array}{rl} \text{ext}V_{sys}^{mat} & : \vec{F}^{sca} \\ \text{int}V_{sys}^{mat} & : \vec{F}^{sca} \end{array}\right\} = \left[\Delta\tilde{\vec{G}}_{sys}^{JF} * \vec{J}^{ES} + \Delta\tilde{\vec{G}}_{sys}^{MF} * \vec{M}^{ES}\right]_{\partial V_{sys}^{mat}} \quad (46)$$

where

$$\tilde{\vec{G}}_{sys}^{JF/MF}(\vec{r},\vec{r}') = \begin{cases} \tilde{\vec{G}}_{sim}^{JF/MF}(\vec{r},\vec{r}') & , \quad (\vec{r}' \in \text{cl}V_{sim}^{mat}) \\ \tilde{\vec{G}}_{mul}^{JF/MF}(\vec{r},\vec{r}') & , \quad (\vec{r}' \in \text{cl}V_{mul}^{mat}) \end{cases} \quad (47)$$

$$\Delta\tilde{\vec{G}}_{sys}^{JF/MF}(\vec{r},\vec{r}') = \begin{cases} \Delta\tilde{\vec{G}}_{sim}^{JF/MF}(\vec{r},\vec{r}') & , \quad (\vec{r}' \in \text{cl}V_{sim}^{mat}) \\ \Delta\tilde{\vec{G}}_{mul}^{JF/MF}(\vec{r},\vec{r}') & , \quad (\vec{r}' \in \text{cl}V_{mul}^{mat}) \end{cases} \quad (48)$$

and

$$\vec{C}^{ES}(\vec{r}') = \begin{cases} \vec{C}_{sim}^{ES}(\vec{r}') & , \quad (\vec{r}' \in \partial V_{sim}^{mat}) \\ \vec{C}_{mul}^{ES}(\vec{r}') & , \quad (\vec{r}' \in \partial V_{mul}^{mat}) \end{cases} \quad (49)$$

in which $C = J, M$.

**Case II: One body contacts with but doesn't submerge the other body**

In this case, the following relationships exist:

$$\partial V_{sys}^{mat} = \partial V_{sim;0}^{mat} \cup \partial V_{mul;0}^{mat} \quad (50.1)$$
$$\text{ext}V_{sys}^{mat} = \text{ext}V_{in}^{mat} \cup \text{ext}V_{out}^{mat} \quad (50.2)$$
$$\text{int}V_{sys}^{mat} = \text{int}V_{sim}^{mat} \cup \text{int}V_{mul}^{mat} \cup \partial V_{\cap}^{mat} \quad (50.3)$$

where the $\partial V_{sim;0}^{mat}$, $\partial V_{mul;0}^{mat}$, $\partial V_{\cap}^{mat}$, $\text{ext}V_{in}^{mat}$, $\text{ext}V_{out}^{mat}$, $\text{int}V_{sim}^{mat}$, and $\text{int}V_{mul}^{mat}$ are shown in Fig. 8 (b).

Based on that $\hat{n}_{sim;\pm} = -\hat{n}_{mul;\pm}$ on $\partial V_{\cap}^{mat}$, the GHPs (29') and (30') are formally specialized to (42) and (43) respectively, and the GFHF (33) is formally specialized to (44).

**Case III: One body submerges the other body**

In this case, the following relationships exist:

$$\partial V_{\cap}^{mat} = \partial V_{sim}^{mat} \quad (51.1)$$
$$\partial V_{sim;0}^{mat} = \varnothing \quad (51.2)$$

and

$$\partial V_{sys}^{mat} = \partial V_{mul;0}^{mat} \quad (52.1)$$

$$\begin{aligned}\text{int}V_{sys}^{mat} &= \text{int}V_{mul}^{mat} \cup \text{int}V_{sim}^{mat} \cup \partial V_{\cap}^{mat} \\ &= \text{int}V_{mul}^{mat} \cup \text{int}V_{sim}^{mat} \cup \partial V_{sim}^{mat} \\ &= \text{int}V_{mul}^{mat} \cup \text{cl}V_{sim}^{mat}\end{aligned} \quad (52.2)$$

where the $\partial V_{sim}^{mat}$, $\partial V_{\cap}^{mat}$, $\partial V_{sys}^{mat}$, $\partial V_{mul;0}^{mat}$, $\text{int}V_{sim}^{mat}$, and $\text{int}V_{mul}^{mat}$ are shown in Fig. 8 (c).

Based on that $\hat{n}_{sim;\pm} = -\hat{n}_{mul;\pm}$ on $\partial V_{\cap}^{mat}$, the GHPs (29') and (30') are formally specialized to (42) and (43) respectively, and the GFHF (33) is formally specialized to (44).

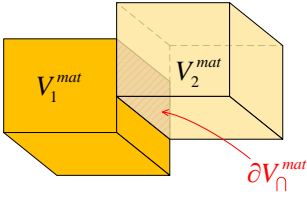

Fig. 9. The EM system constructed by two material bodies.

## V. APPLICATION OF GFHF: TO CONSTRUCT HARRINGTON'S CM OF INHOMOGENEOUS ANISOTROPIC LOSSY MATERIAL SYSTEM

For metallic system, Harrington *et al.* [48] developed a mathematical scheme to construct CM by using SEFIE-MoM (surface electric field integral equation based method of moments). For isotropic material system, Harrington *et al.* constructed some kinds of CM by using VIE-MoM (volume integral equation based MoM) [49] and SIE-MoM (surface integral equation based MoM, also known as PMCHWT-based MoM) [35]. The physical essence of *Harrington's CM* is to construct a series of orthogonal modes which have ability to orthogonalize objective EM power, for example:

• For metallic system, Harrington's SEFIE-based CM [48] orthogonalizes the following objective power:

$$(1/2)\langle \vec{J}^{SL}, \vec{E}^{inc}\rangle_{L^{met}} + (1/2)\langle \vec{J}^{SS}, \vec{E}^{inc}\rangle_{S^{met}\cup\partial V^{met}} \quad (53)$$

where the inner product is defined as $\langle \vec{f}, \vec{g}\rangle_\Omega \triangleq \int_\Omega \vec{f}^* \cdot \vec{g}\, d\Omega$, and $\vec{J}^{SL}$ is the scattering line electric current on metallic line $L^{met}$, and $\vec{J}^{SS}$ is the scattering surface electric current on metallic surface $S^{met}$ and on the boundary of metallic body $\partial V^{met}$.

• For homogeneous or inhomogeneous isotropic material system, Harrington's VIE-based CM [49] orthogonalizes the following objective power:

$$(1/2)\langle \vec{J}^{SV}, \vec{E}^{inc}\rangle_{V^{mat}} + (1/2)\langle \vec{M}^{SV}, \vec{H}^{inc}\rangle_{V^{mat}}. \quad (54)$$

• For homogeneous isotropic material system, Harrington's PMCHWT-based CM [35] orthogonalizes the following power:

$$-(1/2)\langle \vec{J}^{ES}, \vec{E}^{inc}\rangle_{\partial V^{mat}} - (1/2)\langle \vec{M}^{ES}, \vec{H}^{inc}\rangle_{\partial V^{mat}} \quad (55)$$

where the minus signs originate from that the equivalent surface currents in [35] are $\{-\vec{J}^{ES}, -\vec{M}^{ES}\}$.

Recently, [46] proves that the objective powers orthogonalized by VIE-based CM and PMCHWT-based CM are identical to each other, i.e.,

$$\frac{1}{2}\langle \vec{J}^{SV}, \vec{E}^{inc}\rangle_{V^{mat}} + \frac{1}{2}\langle \vec{M}^{SV}, \vec{H}^{inc}\rangle_{V^{mat}} = -\frac{1}{2}\langle \vec{J}^{ES}, \vec{E}^{inc}\rangle_{\partial V^{mat}} - \frac{1}{2}\langle \vec{M}^{ES}, \vec{H}^{inc}\rangle_{\partial V^{mat}} \quad (56)$$

when material system is homogeneous isotropic. In this section, Harrington's CM theory for a simply connected homogeneous isotropic material body [35], [49] is generalized to the EM system which is constructed by multiple inhomogeneous anisotropic lossy material bodies and placed in VACUUM, and the bodies can be either simply or multiply connected. As a typical example, the two-body material system $V_{mat}^{sys}$ shown in Fig. 9 is specifically considered, and the formulations corresponding to any $l$-body material system can be similarly obtained.

The reason to call the CM constructed below as "Harrington's CM" is that the CM orthogonalizes power operator

$$P_{mat\,sys}^{Harrington} = \sum_{i=1}^{2}(1/2)\langle \vec{J}_i^{SV}, \vec{E}^{inc}\rangle_{V_i^{mat}} + (1/2)\langle \vec{M}_i^{SV}, \vec{H}^{inc}\rangle_{V_i^{mat}} \quad (57)$$

by following Harrington's ideas in [49]. In (57), the subscript "$mat\,sys$" is to emphasize that the power operator $P_{mat\,sys}^{Harrington}$ corresponds to material system.

### A. Power characteristic of operator (57)

Similarly to the discussions in our previous paper [46], the power characteristic of $P_{mat\,sys}^{Harrington}$ in (57) can be expressed as follows:

$$P_{mat\,sys}^{Harrington} = P_{mat\,sys}^{sca,rad} + P_{mat\,sys}^{tot,loss,mat} + j\left(P_{mat\,sys}^{sca,sto,field} + P_{mat\,sys}^{tot,sto,mat}\right) \\ - j\omega\left[\langle \Delta\vec{\vec{\mu}}_{mat\,sys} \cdot \vec{H}^{inc}, \vec{H}^{inc}\rangle_{V_{mat}^{sys}} + \text{Re}\{\langle \Delta\vec{\vec{\mu}}_{mat\,sys} \cdot \vec{H}^{sca}, \vec{H}^{inc}\rangle_{V_{mat}^{sys}}\}\right] \quad (58)$$

where

$$P_{mat\,sys}^{sca,rad} = (1/2)\oiint_{S_\infty}\left[\vec{E}^{sca} \times (\vec{H}^{sca})^*\right] \cdot d\vec{S} \quad (59.1)$$

$$P_{mat\,sys}^{sca,sto,field} = 2\omega\left(W_{mat\,sys;m}^{sca,sto,field} - W_{mat\,sys;e}^{sca,sto,field}\right) \quad (59.2)$$

$$P_{mat\,sys}^{tot,loss,mat} = (1/2)\langle \vec{\vec{\sigma}}_{mat\,sys} \cdot \vec{E}^{tot}, \vec{E}^{tot}\rangle_{V_{mat}^{sys}} \quad (59.3)$$

$$P_{mat\,sys}^{tot,sto,mat} = 2\omega\left(W_{mat\,sys;m}^{tot,sto,mat} - W_{mat\,sys;e}^{tot,sto,mat}\right) \quad (59.4)$$

and

$$W_{mat\,sys;m}^{sca,sto,field} = (1/4)\langle \vec{H}^{sca}, \mu_0\vec{H}^{sca}\rangle_{\mathbb{R}^3} \quad (60.1)$$

$$W_{mat\,sys;e}^{sca,sto,field} = (1/4)\langle \varepsilon_0\vec{E}^{sca}, \vec{E}^{sca}\rangle_{\mathbb{R}^3} \quad (60.2)$$

$$W_{mat\,sys;m}^{tot,sto,mat} = (1/4)\langle \vec{H}^{tot}, \Delta\vec{\vec{\mu}}_{mat\,sys} \cdot \vec{H}^{tot}\rangle_{V_{mat}^{sys}} \quad (60.3)$$

$$W_{mat\,sys;e}^{tot,sto,mat} = (1/4)\langle \Delta\vec{\vec{\varepsilon}}_{mat\,sys} \cdot \vec{E}^{tot}, \vec{E}^{tot}\rangle_{V_{mat}^{sys}} \quad (60.4)$$

where $\Delta\vec{\vec{\mu}}_{mat\,sys} = \vec{\vec{\mu}}_{mat\,sys} - \vec{\vec{I}}\mu_0$, and $\Delta\vec{\vec{\varepsilon}}_{mat\,sys} = \vec{\vec{\varepsilon}}_{mat\,sys} - \vec{\vec{I}}\varepsilon_0$, and

$$\vec{\vec{\beta}}_{mat\,sys}(\vec{r}) = \begin{cases}\vec{\vec{\beta}}_1(\vec{r}) &, (\vec{r} \in V_1^{mat}) \\ \vec{\vec{\beta}}_2(\vec{r}) &, (\vec{r} \in V_2^{mat})\end{cases} \quad (61)$$

in which $\beta = \mu, \varepsilon, \sigma$.

### B. Surface formulation of operator (57)

In this section, the surface formulation of Harrington's power operator corresponding to a one-body material system is provided at first, and then the two-body case is discussed.

**One-body case**

By doing some necessary vector operations and utilizing the definition of equivalent surface electric current on $\partial V_i^{mat}$, the following relation can be derived:

$$\left\langle \vec{J}_i^{ES}, \vec{E}^{inc} \right\rangle_{\partial V_i^{mat}} = \oiint_{\partial V_i^{mat}} \left[ \vec{E}^{inc} \times \left( \vec{H}^{tot} \right)^* \right] \cdot d\vec{S} \quad (62)$$

where $i = 1, 2$. In addition,

$$\begin{aligned}
&\oiint_{\partial V_i^{mat}} \left[ \vec{E}^{inc} \times \left( \vec{H}^{tot} \right)^* \right] \cdot d\vec{S} \\
&= \iiint_{V_i^{mat}} \nabla \cdot \left[ \vec{E}^{inc} \times \left( \vec{H}^{tot} \right)^* \right] dV \\
&= \iiint_{V_i^{mat}} \left( \nabla \times \vec{E}^{inc} \right) \cdot \left( \vec{H}^{tot} \right)^* dV - \iiint_{V_i^{mat}} \vec{E}^{inc} \cdot \left( \nabla \times \vec{H}^{tot} \right)^* dV \\
&= \iiint_{V_i^{mat}} \left( -j\omega\mu_0 \vec{H}^{inc} \right) \cdot \left( \vec{H}^{tot} \right)^* dV - \iiint_{V_i^{mat}} \vec{E}^{inc} \cdot \left( j\omega\vec{\varepsilon}_i \cdot \vec{E}^{tot} \right)^* dV \\
&= \iiint_{V_i^{mat}} \left( -j\omega\mu_0 \vec{H}^{inc} \right) \cdot \left( \vec{H}^{tot} \right)^* dV - \iiint_{V_i^{mat}} \vec{E}^{inc} \cdot \left( j\omega\varepsilon_0 \vec{E}^{tot} \right)^* dV \\
&\quad - \iiint_{V_i^{mat}} \vec{E}^{inc} \cdot \left( j\omega\Delta\vec{\varepsilon}_i \cdot \vec{E}^{tot} \right)^* dV \\
&= -j\omega \iiint_{V_i^{mat}} \mu_0 \vec{H}^{inc} \cdot \left( \vec{H}^{tot} \right)^* dV + j\omega \iiint_{V_i^{mat}} \vec{E}^{inc} \cdot \left( \varepsilon_0 \vec{E}^{tot} \right)^* dV \\
&\quad - \iiint_{V_i^{mat}} \vec{E}^{inc} \cdot \left( \vec{J}_i^{SV} \right)^* dV
\end{aligned} \quad (63)$$

In (63), the first equality is due to Gauss' divergence theorem; the second equality is based on that $(\nabla \times \vec{a}) \cdot \vec{b} - \vec{a} \cdot (\nabla \times \vec{b}) = \nabla \cdot (\vec{a} \times \vec{b})$; the third equality is because of the Maxwell's equations of incident field and total field; the forth equality is due to that $\vec{\varepsilon}_i = \vec{I}\varepsilon_0 + \Delta\vec{\varepsilon}_i$; the fifth equality is based on that $\vec{J}_i^{SV} = j\omega\Delta\vec{\varepsilon}_i \cdot \vec{E}^{tot}$ [1], [45]. Then,

$$\left\langle \vec{J}_i^{ES}, \vec{E}^{inc} \right\rangle_{\partial V_i^{mat}} = -j\omega \left\langle \vec{H}^{tot}, \mu_0 \vec{H}^{inc} \right\rangle_{V_i^{mat}} + j\omega \left\langle \varepsilon_0 \vec{E}^{tot}, \vec{E}^{inc} \right\rangle_{V_i^{mat}} - \left\langle \vec{J}_i^{SV}, \vec{E}^{inc} \right\rangle_{V_i^{mat}} \quad (64.1)$$

Similarly, it can be derived that

$$\left\langle \vec{M}_i^{ES}, \vec{H}^{inc} \right\rangle_{\partial V_i^{mat}} = j\omega \left\langle \vec{H}^{tot}, \mu_0 \vec{H}^{inc} \right\rangle_{V_i^{mat}} - j\omega \left\langle \varepsilon_0 \vec{E}^{tot}, \vec{E}^{inc} \right\rangle_{V_i^{mat}} - \left\langle \vec{M}_i^{SV}, \vec{H}^{inc} \right\rangle_{V_i^{mat}} \quad (64.2)$$

The summation of (64.1) and (64.2) gives that

$$\left\langle \vec{J}_i^{SV}, \vec{E}^{inc} \right\rangle_{V_i^{mat}} + \left\langle \vec{M}_i^{SV}, \vec{H}^{inc} \right\rangle_{V_i^{mat}} = -\left\langle \vec{J}_i^{ES}, \vec{E}^{inc} \right\rangle_{\partial V_i^{mat}} - \left\langle \vec{M}_i^{ES}, \vec{H}^{inc} \right\rangle_{\partial V_i^{mat}} \quad (65)$$

(65) is the generalization of the conclusion given in [46].

**Two-body case**

Inserting (65) into (57), the $P_{mat\,sys}^{Harrington}$ can be rewritten as

$$\begin{aligned}
P_{mat\,sys}^{Harrington} &= -\sum_{i=1}^{2} (1/2) \left\langle \vec{J}_i^{ES}, \vec{E}^{inc} \right\rangle_{\partial V_i^{mat}} + (1/2) \left\langle \vec{M}_i^{ES}, \vec{H}^{inc} \right\rangle_{\partial V_i^{mat}} \\
&= -\sum_{i=1}^{2} (1/2) \left\langle \vec{J}_{i0}^{ES} + \vec{J}_{i\cap}^{ES}, \vec{E}^{inc} \right\rangle_{\partial V_i^{mat}} + (1/2) \left\langle \vec{M}_{i0}^{ES} + \vec{M}_{i\cap}^{ES}, \vec{H}^{inc} \right\rangle_{\partial V_i^{mat}} \\
&= -\sum_{i=1}^{2} (1/2) \left\langle \vec{J}_{i0}^{ES}, \vec{E}^{inc} \right\rangle_{\partial V_{i0}^{mat}} + (1/2) \left\langle \vec{M}_{i0}^{ES}, \vec{H}^{inc} \right\rangle_{\partial V_{i0}^{mat}}
\end{aligned} \quad (57')$$

where the second equality is based on (26), and the third equality is based on (28). By utilizing GFHF (44), the surface formulation of power operator $P_{mat\,sys}^{Harrington}$ can be expressed as

$$\begin{aligned}
P_{mat\,sys}^{Harrington} &= \sum_{i=1}^{2} \Big\{ (1/2) \left\langle \vec{J}_{i0}^{ES}, j\omega\mu_0 \mathcal{L}_0 \left( \vec{J}_{10}^{ES} + \vec{J}_{20}^{ES} \right) + \mathcal{K}_0 \left( \vec{M}_{10}^{ES} + \vec{M}_{20}^{ES} \right) \right\rangle_{\partial V_{i0}^{mat}} \\
&\quad + (1/2) \left\langle \vec{M}_{i0}^{ES}, j\omega\varepsilon_0 \mathcal{L}_0 \left( \vec{M}_{10}^{ES} + \vec{M}_{20}^{ES} \right) - \mathcal{K}_0 \left( \vec{J}_{10}^{ES} + \vec{J}_{20}^{ES} \right) \right\rangle_{\partial V_{i0}^{mat}} \Big\}
\end{aligned} \quad (57'')$$

where the $\mathcal{L}_0$ and $\mathcal{K}_0$ are defined as (16). The reason to call (57") as surface formulation is that all arguments in this formulation are surface currents.

### C. Discretization of operator (57")

In this subsection, the operator (57") is transformed from current space to expansion vector space at first, and then the equivalent electric and magnetic currents are related to each other in expansion vector space [46].

**From current space to expansion vector space**

If the currents $\vec{C}_{i0}^{ES}$ and $\vec{C}_{1\cap}^{ES}$ are expanded in terms of proper basis functions as follows:

$$\vec{C}_{i0}^{ES}(\vec{r}) = \sum_{\xi=1}^{\Xi_{i0}^{\vec{C}^{ES}}} a_\xi^{\vec{C}_{i0}^{ES}} \vec{b}_\xi^{\vec{C}_{i0}^{ES}}(\vec{r}) = \bar{\bar{B}}^{\vec{C}_{i0}^{ES}} \cdot \bar{a}^{\vec{C}_{i0}^{ES}} \quad , \quad (\vec{r} \in \partial V_{i0}^{mat}) \quad (66.1)$$

$$\vec{C}_{1\cap}^{ES}(\vec{r}) = \sum_{\xi=1}^{\Xi_{\cap}^{\vec{C}^{ES}}} a_\xi^{\vec{C}_{\cap}^{ES}} \vec{b}_\xi^{\vec{C}_{\cap}^{ES}}(\vec{r}) = \bar{\bar{B}}^{\vec{C}_{\cap}^{ES}} \cdot \bar{a}^{\vec{C}_{\cap}^{ES}} \quad , \quad (\vec{r} \in \partial V_{\cap}^{mat}) \quad (66.2)$$

then

$$\vec{C}_{2\cap}^{ES}(\vec{r}) = -\vec{C}_{1\cap}^{ES}(\vec{r}) = \left( -\bar{\bar{B}}^{\vec{C}_{\cap}^{ES}} \right) \cdot \bar{a}^{\vec{C}_{\cap}^{ES}} \quad , \quad (\vec{r} \in \partial V_{\cap}^{mat}) \quad (66.3)$$

where $C = J, M$, and

$$\bar{\bar{B}}^X = \begin{bmatrix} \vec{b}_1^X & , & \vec{b}_2^X & , & \cdots & , & \vec{b}_{\Xi^X}^X \end{bmatrix} \quad (67.1)$$

$$\bar{a}^X = \begin{bmatrix} a_1^X & , & a_2^X & , & \cdots & , & a_{\Xi^X}^X \end{bmatrix}^T \quad (67.2)$$

for any $X = \vec{C}_{10}^{ES}, \vec{C}_{20}^{ES}, \vec{C}_{1\cap}^{ES}, \vec{C}_{2\cap}^{ES}$.

Inserting (66) into (57"), the objective power $P_{mat\,sys}^{Harrington}$ is discretized to the following matrix form:

$$P_{mat\,sys}^{Harrington} = \left( \bar{a}_{mat\,sys}^{\{\vec{J}_{10}^{ES}, \vec{J}_{\cap}^{ES}, \vec{J}_{20}^{ES}, \vec{M}_{10}^{ES}, \vec{M}_{\cap}^{ES}, \vec{M}_{20}^{ES}\}} \right)^H \cdot \bar{\bar{P}}_{mat\,sys}^{\{\vec{J}_{10}^{ES}, \vec{J}_{\cap}^{ES}, \vec{J}_{20}^{ES}, \vec{M}_{10}^{ES}, \vec{M}_{\cap}^{ES}, \vec{M}_{20}^{ES}\}} \cdot \bar{a}_{mat\,sys}^{\{\vec{J}_{10}^{ES}, \vec{J}_{\cap}^{ES}, \vec{J}_{20}^{ES}, \vec{M}_{10}^{ES}, \vec{M}_{\cap}^{ES}, \vec{M}_{20}^{ES}\}} \quad (68)$$

where

$$\bar{\bar{P}}_{mat\,sys}^{\{\vec{J}_{10}^{ES}, \vec{J}_{\cap}^{ES}, \vec{J}_{20}^{ES}, \vec{M}_{10}^{ES}, \vec{M}_{\cap}^{ES}, \vec{M}_{20}^{ES}\}} = \begin{bmatrix} \bar{\bar{P}}^{\vec{J}_{10}^{ES} \vec{J}_{10}^{ES}} & 0 & \bar{\bar{P}}^{\vec{J}_{10}^{ES} \vec{J}_{20}^{ES}} & \bar{\bar{P}}^{\vec{J}_{10}^{ES} \vec{M}_{10}^{ES}} & 0 & \bar{\bar{P}}^{\vec{J}_{10}^{ES} \vec{M}_{20}^{ES}} \\ 0 & 0 & 0 & 0 & 0 & 0 \\ \bar{\bar{P}}^{\vec{J}_{20}^{ES} \vec{J}_{10}^{ES}} & 0 & \bar{\bar{P}}^{\vec{J}_{20}^{ES} \vec{J}_{20}^{ES}} & \bar{\bar{P}}^{\vec{J}_{20}^{ES} \vec{M}_{10}^{ES}} & 0 & \bar{\bar{P}}^{\vec{J}_{20}^{ES} \vec{M}_{20}^{ES}} \\ \bar{\bar{P}}^{\vec{M}_{10}^{ES} \vec{J}_{10}^{ES}} & 0 & \bar{\bar{P}}^{\vec{M}_{10}^{ES} \vec{J}_{20}^{ES}} & \bar{\bar{P}}^{\vec{M}_{10}^{ES} \vec{M}_{10}^{ES}} & 0 & \bar{\bar{P}}^{\vec{M}_{10}^{ES} \vec{M}_{20}^{ES}} \\ 0 & 0 & 0 & 0 & 0 & 0 \\ \bar{\bar{P}}^{\vec{M}_{20}^{ES} \vec{J}_{10}^{ES}} & 0 & \bar{\bar{P}}^{\vec{M}_{20}^{ES} \vec{J}_{20}^{ES}} & \bar{\bar{P}}^{\vec{M}_{20}^{ES} \vec{M}_{10}^{ES}} & 0 & \bar{\bar{P}}^{\vec{M}_{20}^{ES} \vec{M}_{20}^{ES}} \end{bmatrix} \quad (69.1)$$

$$\bar{a}_{mat\,sys}^{\{\bar{J}_{10}^{ES},\bar{J}_{\cap}^{ES},\bar{J}_{20}^{ES},\bar{M}_{10}^{ES},\bar{M}_{\cap}^{ES},\bar{M}_{20}^{ES}\}} = \begin{bmatrix} \bar{a}^{\bar{J}_{10}^{ES}} \\ \bar{a}^{\bar{J}_{\cap}^{ES}} \\ \bar{a}^{\bar{J}_{20}^{ES}} \\ \bar{a}^{\bar{M}_{10}^{ES}} \\ \bar{a}^{\bar{M}_{\cap}^{ES}} \\ \bar{a}^{\bar{M}_{20}^{ES}} \end{bmatrix} \quad (69.2)$$

in which the elements of various submatrices are as follows:

$$p_{\xi\zeta}^{\bar{J}_{i0}^{ES}\bar{J}_{l0}^{ES}} = j\omega\mu_0\,(1/2)\left\langle \vec{b}_\xi^{\bar{J}_{i0}^{ES}}, \mathcal{L}_0\left(\vec{b}_\zeta^{\bar{J}_{l0}^{ES}}\right)\right\rangle_{\partial V_{i0}^{mat}} \quad (70.1)$$

$$p_{\xi\zeta}^{\bar{J}_{i0}^{ES}\bar{M}_{l0}^{ES}} = (1/2)\left\langle \vec{b}_\xi^{\bar{J}_{i0}^{ES}}, \mathcal{K}_0\left(\vec{b}_\zeta^{\bar{M}_{l0}^{ES}}\right)\right\rangle_{\partial V_{i0-}^{mat}} \quad (70.2)$$

$$p_{\xi\zeta}^{\bar{M}_{i0}^{ES}\bar{M}_{l0}^{ES}} = j\omega\varepsilon_0\,(1/2)\left\langle \vec{b}_\xi^{\bar{M}_{i0}^{ES}}, \mathcal{L}_0\left(\vec{b}_\zeta^{\bar{M}_{l0}^{ES}}\right)\right\rangle_{\partial V_{i0}^{mat}} \quad (70.3)$$

$$p_{\xi\zeta}^{\bar{M}_{i0}^{ES}\bar{J}_{l0}^{ES}} = -(1/2)\left\langle \vec{b}_\xi^{\bar{M}_{i0}^{ES}}, \mathcal{K}_0\left(\vec{b}_\zeta^{\bar{J}_{l0}^{ES}}\right)\right\rangle_{\partial V_{i0-}^{mat}} \quad (70.4)$$

for any $i,l = 1,2$, where the subscript " $-$ " used in integral domain $\partial V_{i0-}^{mat}$ is to emphasize that the integral is done on the internal surface of boundary $\partial V_{i0}^{mat}$.

**To relate equivalent electric and magnetic currents in expansion vector space**

The equivalent electric and magnetic currents on material boundary satisfy the following relations:

$$\partial V_i^{mat}:\; \vec{J}_i^{ES} \times \hat{n}_{\to V_i^{mat}} = \left[\vec{\vec{G}}_{mat\,i}^{JH} * \vec{J}_i^{ES} + \vec{\vec{G}}_{mat\,i}^{MH} * \vec{M}_i^{ES}\right]_{\partial V_i^{mat}}^{tan} \quad (71.1)$$

$$\partial V_i^{mat}:\; \hat{n}_{\to V_i^{mat}} \times \vec{M}_i^{ES} = \left[\vec{\vec{G}}_{mat\,i}^{JE} * \vec{J}_i^{ES} + \vec{\vec{G}}_{mat\,i}^{ME} * \vec{M}_i^{ES}\right]_{\partial V_i^{mat}}^{tan} \quad (71.2)$$

as illustrated in (4) and (45), where the $\vec{\vec{G}}_{mat\,i}^{JH}$, $\vec{\vec{G}}_{mat\,i}^{MH}$, $\vec{\vec{G}}_{mat\,i}^{JE}$, and $\vec{\vec{G}}_{mat\,i}^{ME}$ are the material Green's functions of body $V_i^{mat}$. If (71.1) is tested by $\{\vec{b}_\xi^{\bar{M}_{i0}^{ES}}\}$ and $\{\gamma_i \vec{b}_\xi^{\bar{M}_\cap^{ES}}\}$, then the expansion vectors $\{\bar{a}^{\bar{M}_{i0}^{ES}}, \bar{a}^{\bar{M}_\cap^{ES}}\}$ can be expressed in terms of the expansion vectors $\{\bar{a}^{\bar{J}_{i0}^{ES}}, \bar{a}^{\bar{J}_\cap^{ES}}\}$ as follows:

$$\begin{bmatrix} \bar{a}^{\bar{M}_{i0}^{ES}} \\ \bar{a}^{\bar{M}_\cap^{ES}} \end{bmatrix} = \bar{\bar{T}}_{mat\,sys}^{J_i \to M_i} \cdot \begin{bmatrix} \bar{a}^{\bar{J}_{i0}^{ES}} \\ \bar{a}^{\bar{J}_\cap^{ES}} \end{bmatrix} \quad (72)$$

where

$$\bar{\bar{T}}_{mat\,sys}^{J_i \to M_i} = \begin{bmatrix} \bar{\bar{\Phi}}^{\vec{b}^{\bar{M}_{i0}^{ES}} \bar{M}_{i0}^{ES}} & \bar{\bar{\Phi}}^{\vec{b}^{\bar{M}_{i0}^{ES}} \bar{M}_\cap^{ES}} \\ \bar{\bar{\Phi}}^{\gamma_i \vec{b}^{\bar{M}_\cap^{ES}} \bar{M}_{i0}^{ES}} & \bar{\bar{\Phi}}^{\gamma_i \vec{b}^{\bar{M}_\cap^{ES}} \bar{M}_\cap^{ES}} \end{bmatrix}^{-1} \cdot \begin{bmatrix} \bar{\bar{\Phi}}^{\vec{b}^{\bar{M}_{i0}^{ES}} \bar{J}_{i0}^{ES}} & \bar{\bar{\Phi}}^{\vec{b}^{\bar{M}_{i0}^{ES}} \bar{J}_\cap^{ES}} \\ \bar{\bar{\Phi}}^{\gamma_i \vec{b}^{\bar{M}_\cap^{ES}} \bar{J}_{i0}^{ES}} & \bar{\bar{\Phi}}^{\gamma_i \vec{b}^{\bar{M}_\cap^{ES}} \bar{J}_\cap^{ES}} \end{bmatrix} \quad (73)$$

in which the elements of various submatrices are as follows:

$$\phi_{\xi\zeta}^{\vec{b}^{\bar{M}_{i0}^{ES}} \bar{M}_{i0}^{ES}} = \left\langle \vec{b}_\xi^{\bar{M}_{i0}^{ES}}, \left[\vec{\vec{G}}_{mat\,i}^{MH} * \vec{b}_\zeta^{\bar{M}_{i0}^{ES}}\right]_{\partial V_{i0}^{mat}}\right\rangle_{\partial V_{i0-}^{mat}} \quad (74.1)$$

$$\phi_{\xi\zeta}^{\vec{b}^{\bar{M}_{i0}^{ES}} \bar{M}_\cap^{ES}} = \gamma_i \left\langle \vec{b}_\xi^{\bar{M}_{i0}^{ES}}, \left[\vec{\vec{G}}_{mat\,i}^{MH} * \vec{b}_\zeta^{\bar{M}_\cap^{ES}}\right]_{\partial V_\cap^{mat}}\right\rangle_{\partial V_{i0-}^{mat}} \quad (74.2)$$

$$\phi_{\xi\zeta}^{\gamma_i \vec{b}^{\bar{M}_\cap^{ES}} \bar{M}_{i0}^{ES}} = \gamma_i \left\langle \vec{b}_\xi^{\bar{M}_\cap^{ES}}, \left[\vec{\vec{G}}_{mat\,i}^{MH} * \vec{b}_\zeta^{\bar{M}_{i0}^{ES}}\right]_{\partial V_{i0}^{mat}}\right\rangle_{\partial V_\cap^{mat}} \quad (74.3)$$

$$\phi_{\xi\zeta}^{\gamma_i \vec{b}^{\bar{M}_\cap^{ES}} \bar{M}_\cap^{ES}} = \left\langle \vec{b}_\xi^{\bar{M}_\cap^{ES}}, \left[\vec{\vec{G}}_{mat\,i}^{MH} * \vec{b}_\zeta^{\bar{M}_\cap^{ES}}\right]_{\partial V_\cap^{mat}}\right\rangle_{\partial V_\cap^{mat}} \quad (74.4)$$

$$\phi_{\xi\zeta}^{\vec{b}^{\bar{M}_{i0}^{ES}} \bar{J}_{i0}^{ES}} = \left\langle \vec{b}_\xi^{\bar{M}_{i0}^{ES}}, \vec{b}_\zeta^{\bar{J}_{i0}^{ES}} \times \hat{n}_{\to V_i^{mat}} - \left[\vec{\vec{G}}_{mat\,i}^{JH} * \vec{b}_\zeta^{\bar{J}_{i0}^{ES}}\right]_{\partial V_{i0}^{mat}}\right\rangle_{\partial V_{i0-}^{mat}} \quad (74.5)$$

$$\phi_{\xi\zeta}^{\vec{b}^{\bar{M}_{i0}^{ES}} \bar{J}_\cap^{ES}} = -\gamma_i \left\langle \vec{b}_\xi^{\bar{M}_{i0}^{ES}}, \left[\vec{\vec{G}}_{mat\,i}^{JH} * \vec{b}_\zeta^{\bar{J}_\cap^{ES}}\right]_{\partial V_\cap^{mat}}\right\rangle_{\partial V_{i0-}^{mat}} \quad (74.6)$$

$$\phi_{\xi\zeta}^{\gamma_i \vec{b}^{\bar{M}_\cap^{ES}} \bar{J}_{i0}^{ES}} = -\gamma_i \left\langle \vec{b}_\xi^{\bar{M}_\cap^{ES}}, \left[\vec{\vec{G}}_{mat\,i}^{JH} * \vec{b}_\zeta^{\bar{J}_{i0}^{ES}}\right]_{\partial V_{i0}^{mat}}\right\rangle_{\partial V_\cap^{mat}} \quad (74.7)$$

$$\phi_{\xi\zeta}^{\gamma_i \vec{b}^{\bar{M}_\cap^{ES}} \bar{J}_\cap^{ES}} = \left\langle \vec{b}_\xi^{\bar{M}_\cap^{ES}}, \vec{b}_\zeta^{\bar{J}_\cap^{ES}} \times \hat{n}_{\to V_i^{mat}} - \left[\vec{\vec{G}}_{mat\,i}^{JH} * \vec{b}_\zeta^{\bar{J}_\cap^{ES}}\right]_{\partial V_\cap^{mat}}\right\rangle_{\partial V_\cap^{mat}} \quad (74.8)$$

where $i = 1, 2$, and $\gamma_1 = 1$ and $\gamma_2 = -1$. In fact, the matrix $\bar{\bar{T}}_{mat\,sys}^{J_i \to M_i}$ can be partitioned as follows:

$$\bar{\bar{T}}_{mat\,sys}^{J_i \to M_i} = \begin{bmatrix} \bar{\bar{T}}_{\bar{M}_{i0}^{ES} \bar{J}_{i0}^{ES}}^{J_i \to M_i} & \bar{\bar{T}}_{\bar{M}_{i0}^{ES} \bar{J}_\cap^{ES}}^{J_i \to M_i} \\ \bar{\bar{T}}_{\bar{M}_\cap^{ES} \bar{J}_{i0}^{ES}}^{J_i \to M_i} & \bar{\bar{T}}_{\bar{M}_\cap^{ES} \bar{J}_\cap^{ES}}^{J_i \to M_i} \end{bmatrix} \quad (75)$$

based on the partition way of the vectors in (72). Then, the following transformation from vectors $\{\bar{a}^{\bar{J}_{10}^{ES}}, \bar{a}^{\bar{J}_\cap^{ES}}, \bar{a}^{\bar{J}_{20}^{ES}}\}$ to vectors $\{\bar{a}^{\bar{M}_{10}^{ES}}, \bar{a}^{\bar{M}_\cap^{ES}}, \bar{a}^{\bar{M}_{20}^{ES}}\}$ can be easily established:

$$\begin{bmatrix} \bar{a}^{\bar{M}_{10}^{ES}} \\ \bar{a}^{\bar{M}_\cap^{ES}} \\ \bar{a}^{\bar{M}_{20}^{ES}} \end{bmatrix} = \bar{\bar{T}}_{mat\,sys}^{J \to M} \cdot \begin{bmatrix} \bar{a}^{\bar{J}_{10}^{ES}} \\ \bar{a}^{\bar{J}_\cap^{ES}} \\ \bar{a}^{\bar{J}_{20}^{ES}} \end{bmatrix} \quad (76)$$

where

$$\bar{\bar{T}}_{mat\,sys}^{J \to M} = \begin{bmatrix} \bar{\bar{T}}_{\bar{M}_{10}^{ES} \bar{J}_{10}^{ES}}^{J_1 \to M_1} & \bar{\bar{T}}_{\bar{M}_{10}^{ES} \bar{J}_\cap^{ES}}^{J_1 \to M_1} & 0 \\ \bar{\bar{T}}_{\bar{M}_\cap^{ES} \bar{J}_{10}^{ES}}^{J_1 \to M_1} & \bar{\bar{T}}_{\bar{M}_\cap^{ES} \bar{J}_\cap^{ES}}^{J_1 \to M_1} & 0 \\ 0 & \bar{\bar{T}}_{\bar{M}_{20}^{ES} \bar{J}_\cap^{ES}}^{J_2 \to M_2} & \bar{\bar{T}}_{\bar{M}_{20}^{ES} \bar{J}_{20}^{ES}}^{J_2 \to M_2} \end{bmatrix} \quad (77)$$

or alternatively

$$\bar{\bar{T}}_{mat\,sys}^{J \to M} = \begin{bmatrix} \bar{\bar{T}}_{\bar{M}_{10}^{ES} \bar{J}_{10}^{ES}}^{J_1 \to M_1} & \bar{\bar{T}}_{\bar{M}_{10}^{ES} \bar{J}_\cap^{ES}}^{J_1 \to M_1} & 0 \\ 0 & \bar{\bar{T}}_{\bar{M}_\cap^{ES} \bar{J}_\cap^{ES}}^{J_2 \to M_2} & \bar{\bar{T}}_{\bar{M}_\cap^{ES} \bar{J}_{20}^{ES}}^{J_2 \to M_2} \\ 0 & \bar{\bar{T}}_{\bar{M}_{20}^{ES} \bar{J}_\cap^{ES}}^{J_2 \to M_2} & \bar{\bar{T}}_{\bar{M}_{20}^{ES} \bar{J}_{20}^{ES}}^{J_2 \to M_2} \end{bmatrix} \cdot \quad (78)$$

Inserting (76) into (68), (68) becomes

$$P_{mat\,sys}^{Harrington} = \left(\bar{a}_{mat\,sys}^{\{\bar{J}_{10}^{ES}, \bar{J}_\cap^{ES}, \bar{J}_{20}^{ES}\}}\right)^H \cdot \bar{\bar{P}}_{mat\,sys}^{\{\bar{J}_{10}^{ES}, \bar{J}_\cap^{ES}, \bar{J}_{20}^{ES}\}} \cdot \bar{a}_{mat\,sys}^{\{\bar{J}_{10}^{ES}, \bar{J}_\cap^{ES}, \bar{J}_{20}^{ES}\}} \quad (79)$$

where

$$\bar{\bar{P}}_{mat\,sys}^{\{\bar{J}_{10}^{ES}, \bar{J}_\cap^{ES}, \bar{J}_{20}^{ES}\}} = \begin{bmatrix} \bar{\bar{I}} \\ \bar{\bar{T}}_{mat\,sys}^{J \to M} \end{bmatrix}^H \cdot \bar{\bar{P}}_{mat\,sys}^{\{\bar{J}_{10}^{ES}, \bar{J}_\cap^{ES}, \bar{J}_{20}^{ES}, \bar{M}_{10}^{ES}, \bar{M}_\cap^{ES}, \bar{M}_{20}^{ES}\}} \cdot \begin{bmatrix} \bar{\bar{I}} \\ \bar{\bar{T}}_{mat\,sys}^{J \to M} \end{bmatrix} \quad (80.1)$$

$$\bar{a}_{mat\,sys}^{\{\bar{J}_{10}^{ES}, \bar{J}_\cap^{ES}, \bar{J}_{20}^{ES}\}} = \begin{bmatrix} \bar{a}^{\bar{J}_{10}^{ES}} \\ \bar{a}^{\bar{J}_\cap^{ES}} \\ \bar{a}^{\bar{J}_{20}^{ES}} \end{bmatrix} \quad (80.2)$$

in which $\bar{\bar{I}}$ is the identity matrix whose order is the same as the number of the rows of $\bar{a}_{mat\ sys}^{\{\vec{J}_{10}^{ES},\vec{J}_{\cap}^{ES},\vec{J}_{20}^{ES}\}}$.

Similarly to establishing (76) by testing (71.1) with $\{\vec{b}_\xi^{\vec{M}_{i0}^{ES}}\}$ and $\{\gamma_i \vec{b}_\xi^{\vec{M}_{\cap}^{ES}}\}$, the following transformation from vectors $\{\bar{a}^{\vec{M}_{10}^{ES}},\bar{a}^{\vec{M}_{\cap}^{ES}},\bar{a}^{\vec{M}_{20}^{ES}}\}$ to vectors $\{\bar{a}^{\vec{J}_{10}^{ES}},\bar{a}^{\vec{J}_{\cap}^{ES}},\bar{a}^{\vec{J}_{20}^{ES}}\}$ can be easily established:

$$\begin{bmatrix}\bar{a}^{\vec{J}_{10}^{ES}}\\ \bar{a}^{\vec{J}_{\cap}^{ES}}\\ \bar{a}^{\vec{J}_{20}^{ES}}\end{bmatrix} = \bar{\bar{T}}_{mat\ sys}^{M\to J} \cdot \begin{bmatrix}\bar{a}^{\vec{M}_{10}^{ES}}\\ \bar{a}^{\vec{M}_{\cap}^{ES}}\\ \bar{a}^{\vec{M}_{20}^{ES}}\end{bmatrix} \quad (81)$$

by testing (71.2) with $\{\vec{b}_\xi^{\vec{J}_{i0}^{ES}}\}$ and $\{\gamma_i \vec{b}_\xi^{\vec{J}_{\cap}^{ES}}\}$. Inserting (81) into (68), (68) becomes the following form:

$$P_{mat\ sys}^{Harrington} = \left(\bar{a}_{mat\ sys}^{\{\vec{M}_{10}^{ES},\vec{M}_{\cap}^{ES},\vec{M}_{20}^{ES}\}}\right)^H \cdot \bar{\bar{P}}_{mat\ sys}^{\{\vec{M}_{10}^{ES},\vec{M}_{\cap}^{ES},\vec{M}_{20}^{ES}\}} \cdot \bar{a}_{mat\ sys}^{\{\vec{M}_{10}^{ES},\vec{M}_{\cap}^{ES},\vec{M}_{20}^{ES}\}} \quad (82)$$

where

$$\bar{\bar{P}}_{mat\ sys}^{\{\vec{M}_{10}^{ES},\vec{M}_{\cap}^{ES},\vec{M}_{20}^{ES}\}} = \begin{bmatrix}\bar{\bar{T}}_{mat\ sys}^{M\to J}\\ \bar{\bar{I}}\end{bmatrix}^H \cdot \bar{\bar{P}}_{mat\ sys}^{\{\vec{J}_{10}^{ES},\vec{J}_{\cap}^{ES},\vec{J}_{20}^{ES},\vec{M}_{10}^{ES},\vec{M}_{\cap}^{ES},\vec{M}_{20}^{ES}\}} \cdot \begin{bmatrix}\bar{\bar{T}}_{mat\ sys}^{M\to J}\\ \bar{\bar{I}}\end{bmatrix} \quad (83.1)$$

$$\bar{a}_{mat\ sys}^{\{\vec{M}_{10}^{ES},\vec{M}_{\cap}^{ES},\vec{M}_{20}^{ES}\}} = \begin{bmatrix}\bar{a}^{\vec{M}_{10}^{ES}}\\ \bar{a}^{\vec{M}_{\cap}^{ES}}\\ \bar{a}^{\vec{M}_{20}^{ES}}\end{bmatrix}. \quad (83.2)$$

For the convenience of the following discussions, (79) and (82) are uniformly written as follows:

$$P_{mat\ sys}^{Harrington} = \left(\bar{a}_{mat\ sys}^{\{\vec{C}_{10}^{ES},\vec{C}_{\cap}^{ES},\vec{C}_{20}^{ES}\}}\right)^H \cdot \bar{\bar{P}}_{mat\ sys}^{\{\vec{C}_{10}^{ES},\vec{C}_{\cap}^{ES},\vec{C}_{20}^{ES}\}} \cdot \bar{a}_{mat\ sys}^{\{\vec{C}_{10}^{ES},\vec{C}_{\cap}^{ES},\vec{C}_{20}^{ES}\}} \quad (84)$$

where $C = J, M$.

### D. Harrington's CM orthogonalizing operator (57)

The power matrix $\bar{\bar{P}}^{\{\vec{C}_{10}^{ES},\vec{C}_{\cap}^{ES},\vec{C}_{20}^{ES}\}}$ can be decomposed as

$$\bar{\bar{P}}_{mat\ sys}^{\{\vec{C}_{10}^{ES},\vec{C}_{\cap}^{ES},\vec{C}_{20}^{ES}\}} = \bar{\bar{P}}_{mat\ sys;+}^{\{\vec{C}_{10}^{ES},\vec{C}_{\cap}^{ES},\vec{C}_{20}^{ES}\}} + j\bar{\bar{P}}_{mat\ sys;-}^{\{\vec{C}_{10}^{ES},\vec{C}_{\cap}^{ES},\vec{C}_{20}^{ES}\}} \quad (85)$$

where [46]

$$\bar{\bar{P}}_{mat\ sys;+}^{\{\vec{C}_{10}^{ES},\vec{C}_{\cap}^{ES},\vec{C}_{20}^{ES}\}} = \frac{1}{2}\left[\bar{\bar{P}}_{mat\ sys}^{\{\vec{C}_{10}^{ES},\vec{C}_{\cap}^{ES},\vec{C}_{20}^{ES}\}} + \left(\bar{\bar{P}}_{mat\ sys}^{\{\vec{C}_{10}^{ES},\vec{C}_{\cap}^{ES},\vec{C}_{20}^{ES}\}}\right)^H\right] \quad (86.1)$$

$$\bar{\bar{P}}_{mat\ sys;-}^{\{\vec{C}_{10}^{ES},\vec{C}_{\cap}^{ES},\vec{C}_{20}^{ES}\}} = \frac{1}{2j}\left[\bar{\bar{P}}_{mat\ sys}^{\{\vec{C}_{10}^{ES},\vec{C}_{\cap}^{ES},\vec{C}_{20}^{ES}\}} - \left(\bar{\bar{P}}_{mat\ sys}^{\{\vec{C}_{10}^{ES},\vec{C}_{\cap}^{ES},\vec{C}_{20}^{ES}\}}\right)^H\right]. \quad (86.2)$$

Based on Harrington's classical method [35], [48], [49], the CM can be obtained by solving characteristic equation

$$\bar{\bar{P}}_{mat\ sys;-}^{\{\vec{C}_{10}^{ES},\vec{C}_{\cap}^{ES},\vec{C}_{20}^{ES}\}} \cdot \bar{a}_{mat\ sys;\xi}^{\{\vec{C}_{10}^{ES},\vec{C}_{\cap}^{ES},\vec{C}_{20}^{ES}\}} = \lambda_{mat\ sys;\xi} \bar{\bar{P}}_{mat\ sys;+}^{\{\vec{C}_{10}^{ES},\vec{C}_{\cap}^{ES},\vec{C}_{20}^{ES}\}} \cdot \bar{a}_{mat\ sys;\xi}^{\{\vec{C}_{10}^{ES},\vec{C}_{\cap}^{ES},\vec{C}_{20}^{ES}\}}. \quad (87)$$

In addition, the electromagnetic-power-based (EMP-based) CMT for the inhomogeneous anisotropic material system can be easily established by employing the GFHFs obtained in this paper and the formulations provided in paper [51], and it will not be repeated here.

## VI. CONCLUSIONS

In this paper, the EM diffraction integral formulations in homogeneous isotropic media are generalized to inhomogeneous anisotropic lossy media. Then the traditional HP, ET, and FHF of a single simply connected homogeneous isotropic material body in homogeneous isotropic environment are generalized to the EM system which is constructed by several simply or multiply connected inhomogeneous anisotropic lossy material bodies (the different bodies can either contact or be non-contact with each other) and placed in an inhomogeneous anisotropic lossy environment; the traditional FHF of external scattering field and internal total field are generalized to the internal incident field and internal scattering field, and the equivalent surface currents used to express these fields are the same. The generalized versions of HP, ET, and FHF satisfy so-called topological additivity, i.e., the GHP/GET/GFHF of whole EM system equals to the summation of the GHP/GET/GFHF corresponding to all sub-systems.

The relationships among HP, ET, and FHF are studied, and it is found out that the mathematical formulation of HP and ET are essentially equivalent to each other; the FHF is not the mathematical expression of HP, and it is only the mathematical expression of SEP; HP is a special SEP, and SEP is not necessarily HP; HP can be viewed as physical equivalence principle, because it simultaneously satisfies the action at a distance, the law of causality, and the principle of superposition. Based on these observations, the reason leading to the backward wave problem of FHF is clearly explained.

Compared with the HP and ET, the FHF has not a clearer physical meaning, but it doesn't imply that the FHF is useless. The values of FHF are mainly manifested in that various EM fields are uniformly expressed in terms of an identical set of currents, and this feature is very valuable for many engineering applications as exhibited in this paper.

## APPENDIX A: INTEGRAL EXPRESSIONS OF THE FIELDS IN AN INHOMOGENEOUS ANISOTROPIC OPEN DOMAIN $\Omega$

In this appendix A, some integral expressions for the fields in inhomogeneous anisotropic environment are derived, and the expressions are based on EM dyadic Green's functions according to Prof. Tai's observation "*... the most compact formulation appears to be the one based on the dyadic Green's function pertaining to the vector wave equation for $\vec{E}$ and $\vec{H}$ ...*" [28].

In any open domain $\Omega$ whose material parameters are $\{\bar{\bar{\varepsilon}}_\Omega, \bar{\bar{\mu}}_\Omega\}$, it is supposed that the EM fields $\{\vec{E}_\Omega, \vec{H}_\Omega\}$ and currents $\{\vec{J}_\Omega, \vec{M}_\Omega\}$ satisfy the following Maxwell's equations:

$$\begin{aligned}\nabla \times \vec{H}_\Omega(\vec{r}) &= \vec{J}_\Omega(\vec{r}) + j\omega \ddot{\varepsilon}_\Omega(\vec{r}) \cdot \vec{E}_\Omega(\vec{r}) \\ \nabla \times \vec{E}_\Omega(\vec{r}) &= -\vec{M}_\Omega(\vec{r}) - j\omega \ddot{\mu}_\Omega(\vec{r}) \cdot \vec{H}_\Omega(\vec{r})\end{aligned} \quad \text{(A-1)}$$

for any $\vec{r} \in \Omega$, where the terminology "open domain" means that $\Omega = \text{int}\Omega$ [38], and the subscripts "$\Omega$" used in various quantities mean that these quantities distribute on domain $\Omega$.

Various EM dyadic Green's functions on domain $\Omega$ are defined as follows: [1], [37], [45]

$$\begin{aligned}\nabla \times \ddot{G}_\Omega^{JH}(\vec{r},\vec{r}') &= \vec{I}\delta(\vec{r}-\vec{r}') + j\omega \ddot{\varepsilon}_\Omega(\vec{r}) \cdot \ddot{G}_\Omega^{JE}(\vec{r},\vec{r}') \\ \nabla \times \ddot{G}_\Omega^{JE}(\vec{r},\vec{r}') &= -j\omega \ddot{\mu}_\Omega(\vec{r}) \cdot \ddot{G}_\Omega^{JH}(\vec{r},\vec{r}')\end{aligned} \quad \text{(A-2.1)}$$

for the dyadic Green's functions corresponding to electric-type unity dyadic point source, and

$$\begin{aligned}\nabla \times \ddot{G}_\Omega^{MH}(\vec{r},\vec{r}') &= j\omega \ddot{\varepsilon}_\Omega(\vec{r}) \cdot \ddot{G}_\Omega^{ME}(\vec{r},\vec{r}') \\ \nabla \times \ddot{G}_\Omega^{ME}(\vec{r},\vec{r}') &= -\vec{I}\delta(\vec{r}-\vec{r}') - j\omega \ddot{\mu}_\Omega(\vec{r}) \cdot \ddot{G}_\Omega^{MH}(\vec{r},\vec{r}')\end{aligned} \quad \text{(A-2.2)}$$

for the dyadic Green's functions corresponding to magnetic-type unity dyadic point source. In (A-2), the $\vec{I}$ is identity dyad, and the $\delta(\vec{r}-\vec{r}')$ is Dirac delta function, and $\vec{r}, \vec{r}' \in \Omega$.

If $\ddot{\alpha}$ is a two-order complex symmetrical tensor, its inverse $\ddot{\alpha}^{-1}$ is also symmetrical because of the following observation:

$$\begin{aligned}(\ddot{\alpha}^{-1})^T \cdot \ddot{\alpha}^T &= (\ddot{\alpha} \cdot \ddot{\alpha}^{-1})^T = \vec{I}^T = \vec{I} \\ \Rightarrow (\ddot{\alpha}^{-1})^T &= (\ddot{\alpha}^T)^{-1} = \ddot{\alpha}^{-1}\end{aligned} \quad \text{(A-3)}$$

Based on the identity $\nabla \cdot (\vec{a} \times \vec{b}) = (\nabla \times \vec{a}) \cdot \vec{b} - \vec{a} \cdot (\nabla \times \vec{b})$ [37] and the symmetry of $\ddot{\alpha}^{-1}$, the following (A-4) can be obtained:

$$\begin{aligned}&\nabla \cdot \left\{ \vec{P} \times \left[\ddot{\alpha}^{-1} \cdot (\nabla \times \ddot{Q})\right] + \left[\ddot{\alpha}^{-1} \cdot (\nabla \times \vec{P})\right] \times \ddot{Q} \right\} \\ &= \left\{\nabla \times \left[\ddot{\alpha}^{-1} \cdot (\nabla \times \vec{P})\right]\right\} \cdot \ddot{Q} - \vec{P} \cdot \left\{\nabla \times \left[\ddot{\alpha}^{-1} \cdot (\nabla \times \ddot{Q})\right]\right\}\end{aligned} \quad \text{(A-4)}$$

Appling divergence theorem to (A-4), the following *generalized vector-dyadic Green's second theorem* can be derived:

$$\begin{aligned}&\iiint_\Omega \left(\vec{P} \cdot \left\{\nabla \times \left[\ddot{\alpha}^{-1} \cdot (\nabla \times \ddot{Q})\right]\right\} - \left\{\nabla \times \left[\ddot{\alpha}^{-1} \cdot (\nabla \times \vec{P})\right]\right\} \cdot \ddot{Q}\right)d\Omega \\ &= \oiint_{\partial\Omega} \hat{n}_{\to\Omega} \cdot \left\{\vec{P} \times \left[\ddot{\alpha}^{-1} \cdot (\nabla \times \ddot{Q})\right] + \left[\ddot{\alpha}^{-1} \cdot (\nabla \times \vec{P})\right] \times \ddot{Q}\right\}dS\end{aligned} \quad \text{(A-5)}$$

In (A-5), $\hat{n}_{\to\Omega}$ is the unity normal vector of boundary $\partial\Omega$, and it points to the interior of domain $\Omega$.

Inserting $\vec{P} = \vec{E}_\Omega(\vec{r})$ and $\ddot{Q} = \ddot{G}_\Omega^{JE}(\vec{r},\vec{r}')$ and $\ddot{\alpha} = \ddot{\mu}_\Omega(\vec{r})$ into (A-5), and restricting that there doesn't exist surface magnetization magnetic current on $\partial\Omega$ (the reasonability of this restriction will be explained in Appendix C), the following relation can be derived:

$$\begin{aligned}&-j\omega\iiint_\Omega \vec{E}_\Omega(\vec{r}) \cdot \vec{I}\delta(\vec{r}-\vec{r}')d\Omega \\ &+j\omega\iiint_\Omega \vec{J}_\Omega(\vec{r}) \cdot \ddot{G}_\Omega^{JE}(\vec{r},\vec{r}')d\Omega \\ &+\omega^2 \iiint_\Omega \vec{E}_\Omega(\vec{r}) \cdot \left[\ddot{\varepsilon}_\Omega(\vec{r}) \cdot \ddot{G}_\Omega^{JE}(\vec{r},\vec{r}')\right]d\Omega \\ &-\omega^2 \iiint_\Omega \left[\ddot{\varepsilon}_\Omega(\vec{r}) \cdot \vec{E}_\Omega(\vec{r})\right] \cdot \ddot{G}_\Omega^{JE}(\vec{r},\vec{r}')d\Omega \\ &+\iiint_\Omega \left\{\nabla \times \left[\ddot{\mu}_\Omega^{-1}(\vec{r}) \cdot \vec{M}_\Omega(\vec{r})\right]\right\} \cdot \ddot{G}_\Omega^{JE}(\vec{r},\vec{r}')d\Omega \\ &= -j\omega \oiint_{\partial\Omega} \hat{n}_{\to\Omega} \cdot \left[\left\{\ddot{\mu}_\Omega^{-1}(\vec{r}) \cdot \left[\ddot{\mu}_\Omega(\vec{r}) \cdot \vec{H}_\Omega(\vec{r})\right]\right\} \times \ddot{G}_\Omega^{JE}(\vec{r},\vec{r}')\right]dS \\ &-j\omega \oiint_{\partial\Omega} \hat{n}_{\to\Omega} \cdot \left[\vec{E}_\Omega(\vec{r}) \times \left\{\ddot{\mu}_\Omega^{-1}(\vec{r}) \cdot \left[\ddot{\mu}_\Omega(\vec{r}) \cdot \ddot{G}_\Omega^{JH}(\vec{r},\vec{r}')\right]\right\}\right]dS \\ &- \oiint_{\partial\Omega} \hat{n}_{\to\Omega} \cdot \left\{\left[\ddot{\mu}_\Omega^{-1}(\vec{r}) \cdot \vec{M}_\Omega(\vec{r})\right] \times \ddot{G}_\Omega^{JE}(\vec{r},\vec{r}')\right\}dS\end{aligned} \quad \text{(A-6)}$$

where the integrals on boundary surface $\partial\Omega$ are defined as $\oiint_{\partial\Omega} A(\vec{r},\vec{r}')dS \triangleq \lim_{\Gamma \to \partial\Omega} \oiint_\Gamma A(\vec{r},\vec{r}')dS$, and $\Gamma \subset \Omega$. The last term in the left-hand side of (A-6) can be rewritten as follows:

$$\begin{aligned}&\iiint_\Omega \left\{\nabla \times \left[\ddot{\mu}_\Omega^{-1}(\vec{r}) \cdot \vec{M}_\Omega(\vec{r})\right]\right\} \cdot \ddot{G}_\Omega^{JE}(\vec{r},\vec{r}')d\Omega \\ &= -\oiint_{\partial\Omega} \hat{n}_{\to\Omega} \cdot \left\{\left[\ddot{\mu}_\Omega^{-1}(\vec{r}) \cdot \vec{M}_\Omega(\vec{r})\right] \times \ddot{G}_\Omega^{JE}(\vec{r},\vec{r}')\right\}dS \\ &+ \iiint_\Omega \left[\ddot{\mu}_\Omega^{-1}(\vec{r}) \cdot \vec{M}_\Omega(\vec{r})\right] \cdot \left[\nabla \times \ddot{G}_\Omega^{JE}(\vec{r},\vec{r}')\right]d\Omega \\ &= -\oiint_{\partial\Omega} \hat{n}_{\to\Omega} \cdot \left\{\left[\ddot{\mu}_\Omega^{-1}(\vec{r}) \cdot \vec{M}_\Omega(\vec{r})\right] \times \ddot{G}_\Omega^{JE}(\vec{r},\vec{r}')\right\}dS \\ &-j\omega \iiint_\Omega \left[\ddot{\mu}_\Omega^{-1}(\vec{r}) \cdot \vec{M}_\Omega(\vec{r})\right] \cdot \left[\ddot{\mu}_\Omega(\vec{r}) \cdot \ddot{G}_\Omega^{JH}(\vec{r},\vec{r}')\right]d\Omega\end{aligned} \quad \text{(A-7)}$$

where the first equality is based on the identity $\nabla \cdot (\vec{a} \times \vec{b}) = (\nabla \times \vec{a}) \cdot \vec{b} - \vec{a} \cdot (\nabla \times \vec{b})$ and Gauss' divergence theorem [37], and the second equality is based on (A-2.1). Because the dot product of two dyads satisfies the associative property [37] and the $\ddot{\varepsilon}_\Omega$ is symmetrical, then

$$\begin{aligned}\vec{E}_\Omega(\vec{r}) \cdot \left[\ddot{\varepsilon}_\Omega(\vec{r}) \cdot \ddot{G}_\Omega^{JE}(\vec{r},\vec{r}')\right] &= \left[\vec{E}_\Omega(\vec{r}) \cdot \ddot{\varepsilon}_\Omega(\vec{r})\right] \cdot \ddot{G}_\Omega^{JE}(\vec{r},\vec{r}') \\ &= \left[\ddot{\varepsilon}_\Omega(\vec{r}) \cdot \vec{E}_\Omega(\vec{r})\right] \cdot \ddot{G}_\Omega^{JE}(\vec{r},\vec{r}')\end{aligned} \quad \text{(A-8.1)}$$

Similarly, it can be derived that

$$\left[\ddot{\mu}_\Omega^{-1}(\vec{r}) \cdot \vec{M}_\Omega(\vec{r})\right] \cdot \left[\ddot{\mu}_\Omega(\vec{r}) \cdot \ddot{G}_\Omega^{JH}(\vec{r},\vec{r}')\right] = \vec{M}_\Omega(\vec{r}) \cdot \ddot{G}_\Omega^{JH}(\vec{r},\vec{r}') \quad \text{(A-8.2)}$$

$$\left\{\ddot{\mu}_\Omega^{-1}(\vec{r}) \cdot \left[\ddot{\mu}_\Omega(\vec{r}) \cdot \vec{H}_\Omega(\vec{r})\right]\right\} \times \ddot{G}_\Omega^{JE}(\vec{r},\vec{r}') = \vec{H}_\Omega(\vec{r}) \times \ddot{G}_\Omega^{JE}(\vec{r},\vec{r}') \quad \text{(A-8.3)}$$

$$\vec{E}_\Omega(\vec{r}) \times \left\{\ddot{\mu}_\Omega^{-1}(\vec{r}) \cdot \left[\ddot{\mu}_\Omega(\vec{r}) \cdot \ddot{G}_\Omega^{JH}(\vec{r},\vec{r}')\right]\right\} = \vec{E}_\Omega(\vec{r}) \times \ddot{G}_\Omega^{JH}(\vec{r},\vec{r}') \quad \text{(A-8.4)}$$

based on the symmetry of $\ddot{\mu}_\Omega$ and $\ddot{\mu}_\Omega^{-1}$. Inserting (A-7) and (A-8) into (A-6) and utilizing the property that $\vec{E}_\Omega(\vec{r}') = \iiint_\Omega \vec{E}_\Omega(\vec{r}) \cdot \vec{I}\delta(\vec{r}-\vec{r}')d\Omega$, (A-6) can be simplified as

$$\begin{aligned}\vec{E}_\Omega(\vec{r}') = &\iiint_\Omega \vec{J}_\Omega(\vec{r}) \cdot \ddot{G}_\Omega^{JE}(\vec{r},\vec{r}')d\Omega \\ &- \iiint_\Omega \vec{M}_\Omega(\vec{r}) \cdot \ddot{G}_\Omega^{JH}(\vec{r},\vec{r}')d\Omega \\ &+ \oiint_{\partial\Omega} \hat{n}_{\to\Omega} \cdot \left[\vec{H}_\Omega(\vec{r}) \times \ddot{G}_\Omega^{JE}(\vec{r},\vec{r}')\right]dS \\ &+ \oiint_{\partial\Omega} \hat{n}_{\to\Omega} \cdot \left[\vec{E}_\Omega(\vec{r}) \times \ddot{G}_\Omega^{JH}(\vec{r},\vec{r}')\right]dS\end{aligned} \quad \text{(A-9)}$$

Based on the identity $\vec{a}\cdot(\vec{b}\times\vec{c})=\vec{b}\cdot(\vec{c}\times\vec{a})=\vec{c}\cdot(\vec{a}\times\vec{b})$, (A-9) can be further rewritten as

$$\begin{aligned}\vec{E}_\Omega(\vec{r}') = &\iiint_\Omega \vec{J}_\Omega(\vec{r})\cdot\vec{\vec{G}}_\Omega^{JE}(\vec{r},\vec{r}')d\Omega \\ &- \iiint_\Omega \vec{M}_\Omega(\vec{r})\cdot\vec{\vec{G}}_\Omega^{JH}(\vec{r},\vec{r}')d\Omega \\ &+ \oiint_{\partial\Omega}\left[\hat{n}_{\to\Omega}\times\vec{H}_\Omega(\vec{r})\right]\cdot\vec{\vec{G}}_\Omega^{JE}(\vec{r},\vec{r}')dS \\ &- \oiint_{\partial\Omega}\left[\vec{E}_\Omega(\vec{r})\times\hat{n}_{\to\Omega}\right]\cdot\vec{\vec{G}}_\Omega^{JH}(\vec{r},\vec{r}')dS\end{aligned} \quad \text{(A-10)}$$

Interchanging the position vectors $\vec{r}$ and $\vec{r}'$ in (A-10), the following integral formulation of $\vec{E}_\Omega$ is derived:

$$\begin{aligned}\vec{E}_\Omega(\vec{r}) = &\iiint_\Omega \vec{J}_\Omega(\vec{r}')\cdot\vec{\vec{G}}_\Omega^{JE}(\vec{r}',\vec{r})d\Omega' \\ &- \iiint_\Omega \vec{M}_\Omega(\vec{r}')\cdot\vec{\vec{G}}_\Omega^{JH}(\vec{r}',\vec{r})d\Omega' \\ &+ \oiint_{\partial\Omega}\left[\hat{n}_{\to\Omega}\times\vec{H}_\Omega(\vec{r}')\right]\cdot\vec{\vec{G}}_\Omega^{JE}(\vec{r}',\vec{r})dS' \\ &- \oiint_{\partial\Omega}\left[\vec{E}_\Omega(\vec{r}')\times\hat{n}_{\to\Omega}\right]\cdot\vec{\vec{G}}_\Omega^{JH}(\vec{r}',\vec{r})dS'\end{aligned} \quad \text{(A-11)}$$

for any $\vec{r}\in\Omega$. If the following boundary currents are introduced:

$$\vec{J}_{\partial\Omega}(\vec{r}) \triangleq \hat{n}_{\to\Omega}(\vec{r})\times\left[\vec{H}_\Omega(\vec{r}')\right]_{\vec{r}'\to\vec{r}}, \quad (\vec{r}\in\partial\Omega) \quad \text{(A-12.1)}$$
$$\vec{M}_{\partial\Omega}(\vec{r}) \triangleq \left[\vec{E}_\Omega(\vec{r}')\right]_{\vec{r}'\to\vec{r}}\times\hat{n}_{\to\Omega}(\vec{r}), \quad (\vec{r}\in\partial\Omega) \quad \text{(A-12.2)}$$

where $\vec{r}'\in\text{int}\,\Omega$ and $\vec{r}'$ tends to $\vec{r}$, then (A-11) can be rewritten as follows:

$$\begin{aligned}\vec{E}_\Omega(\vec{r}) = &\iiint_\Omega \vec{J}_\Omega(\vec{r}')\cdot\vec{\vec{G}}_\Omega^{JE}(\vec{r}',\vec{r})d\Omega' \\ &- \iiint_\Omega \vec{M}_\Omega(\vec{r}')\cdot\vec{\vec{G}}_\Omega^{JH}(\vec{r}',\vec{r})d\Omega' \\ &+ \oiint_{\partial\Omega}\vec{J}_{\partial\Omega}(\vec{r}')\cdot\vec{\vec{G}}_\Omega^{JE}(\vec{r}',\vec{r})dS' \\ &- \oiint_{\partial\Omega}\vec{M}_{\partial\Omega}(\vec{r}')\cdot\vec{\vec{G}}_\Omega^{JH}(\vec{r}',\vec{r})dS'\end{aligned} \quad \text{(A-13)}$$

If $\vec{P}=\vec{H}_\Omega(\vec{r})$ and $\vec{Q}=\vec{\vec{G}}_\Omega^{MH}(\vec{r},\vec{r}')$ and $\vec{\vec{\alpha}}=\vec{\vec{\varepsilon}}_\Omega(\vec{r})$ are inserted into (A-5), the following integral expression for the magnetic field $\vec{H}_\Omega$ at any position $\vec{r}$ in $\Omega$ can be obtained similarly:

$$\begin{aligned}\vec{H}_\Omega(\vec{r}) = &- \iiint_\Omega \vec{J}_\Omega(\vec{r}')\cdot\vec{\vec{G}}_\Omega^{ME}(\vec{r}',\vec{r})d\Omega' \\ &+ \iiint_\Omega \vec{M}_\Omega(\vec{r}')\cdot\vec{\vec{G}}_\Omega^{MH}(\vec{r}',\vec{r})d\Omega' \\ &- \oiint_{\partial\Omega}\vec{J}_{\partial\Omega}(\vec{r}')\cdot\vec{\vec{G}}_\Omega^{ME}(\vec{r}',\vec{r})dS' \\ &+ \oiint_{\partial\Omega}\vec{M}_{\partial\Omega}(\vec{r}')\cdot\vec{\vec{G}}_\Omega^{MH}(\vec{r}',\vec{r})dS'\end{aligned} \quad \text{(A-14)}$$

for any $\vec{r}\in\Omega$.

APPENDIX B: THE SYMMETRY OF THE DYADIC GREEN'S FUNCTIONS CORRESPONDING TO AN INHOMOGENEOUS ANISOTROPIC OPEN DOMAIN $\Omega$

Based on (A-2.2), it can be concluded that the magnetic-type identity vector point source $\hat{\xi}\delta(\vec{r}-\vec{r}'')$ and its fields $\{\vec{\vec{G}}_{\Omega;\xi}^{ME}(\vec{r},\vec{r}''),\vec{\vec{G}}_{\Omega;\xi}^{MH}(\vec{r},\vec{r}'')\}$ satisfy the following Maxwell's equations:

$$\begin{aligned}\nabla\times\vec{G}_{\Omega;\xi}^{MH}(\vec{r},\vec{r}'') &= j\omega\vec{\vec{\varepsilon}}_\Omega(\vec{r})\cdot\vec{G}_{\Omega;\xi}^{ME}(\vec{r},\vec{r}'') \\ \nabla\times\vec{G}_{\Omega;\xi}^{ME}(\vec{r},\vec{r}'') &= -\hat{\xi}\delta(\vec{r}-\vec{r}'')-j\omega\vec{\vec{\mu}}_\Omega(\vec{r})\cdot\vec{G}_{\Omega;\xi}^{MH}(\vec{r},\vec{r}'')\end{aligned} \quad \text{(B-1)}$$

in which $\xi=x,y,z$, and

$$\vec{G}_{\Omega;\xi}^{MF}(\vec{r},\vec{r}'') = \hat{x}G_{\Omega;x\xi}^{MF}(\vec{r},\vec{r}'')+\hat{y}G_{\Omega;y\xi}^{MF}(\vec{r},\vec{r}'')+\hat{z}G_{\Omega;z\xi}^{MF}(\vec{r},\vec{r}'') \quad \text{(B-2)}$$

where $F=E,H$.

If the currents in (A-13) are as follows:

$$\{\vec{J}_\Omega(\vec{r}),\vec{J}_{\partial\Omega}(\vec{r})\} = \{0\} \quad \text{(B-3.1)}$$
$$\{\vec{M}_\Omega(\vec{r}),\vec{M}_{\partial\Omega}(\vec{r})\} = \{\hat{\xi}\delta(\vec{r}-\vec{r}'')\} \quad \text{(B-3.2)}$$

then (A-13) gives that

$$\begin{aligned}\vec{G}_{\Omega;\xi}^{ME}(\vec{r},\vec{r}'') &= - \int_{\Omega\cup\partial\Omega}\left[\hat{\xi}\delta(\vec{r}'-\vec{r}'')\right]\cdot\vec{\vec{G}}_\Omega^{JH}(\vec{r}',\vec{r})d\Pi' \\ &= - \int_{\Omega\cup\partial\Omega}\delta(\vec{r}'-\vec{r}'')\vec{G}_{\Omega;\xi\cdot}^{JH}(\vec{r}',\vec{r})d\Pi' \\ &= - \vec{G}_{\Omega;\xi\cdot}^{JH}(\vec{r}'',\vec{r})\end{aligned} \quad \text{(B-4)}$$

where

$$\vec{G}_{\Omega;\xi\cdot}^{JH}(\vec{r}'',\vec{r}) = \hat{x}G_{\Omega;\xi x}^{JH}(\vec{r}'',\vec{r})+\hat{y}G_{\Omega;\xi y}^{JH}(\vec{r}'',\vec{r})+\hat{z}G_{\Omega;\xi z}^{JH}(\vec{r}'',\vec{r}). \quad \text{(B-5)}$$

In fact, (B-4) is equivalent to saying that

$$\vec{\vec{G}}_\Omega^{ME}(\vec{r},\vec{r}') = -\left[\vec{\vec{G}}_\Omega^{JH}(\vec{r}',\vec{r})\right]^T \quad \text{(B-6.1)}$$
$$\vec{\vec{G}}_\Omega^{JH}(\vec{r},\vec{r}') = -\left[\vec{\vec{G}}_\Omega^{ME}(\vec{r}',\vec{r})\right]^T \quad \text{(B-6.2)}$$

where the superscript "$T$" is the transpose of a dyad.

Similarly to (A-5), the following *generalized vector-vector Green's second theorem* exists:

$$\begin{aligned}&\iiint_\Omega\left(\vec{P}\cdot\left\{\nabla\times\left[\vec{\vec{\alpha}}^{-1}\cdot(\nabla\times\vec{Q})\right]\right\}-\left\{\nabla\times\left[\vec{\vec{\alpha}}^{-1}\cdot(\nabla\times\vec{P})\right]\right\}\cdot\vec{Q}\right)d\Omega \\ &= \oiint_{\partial\Omega}\hat{n}_{\to\Omega}\cdot\left\{\vec{P}\times\left[\vec{\vec{\alpha}}^{-1}\cdot(\nabla\times\vec{Q})\right]+\left[\vec{\vec{\alpha}}^{-1}\cdot(\nabla\times\vec{P})\right]\times\vec{Q}\right\}dS\end{aligned}. \quad \text{(B-7)}$$

It is supposed that the EM fields $\{\vec{E}_\Omega^{\vec{J}_\Omega^a},\vec{H}_\Omega^{\vec{J}_\Omega^a}\}/\{\vec{E}_\Omega^{\vec{J}_\Omega^b},\vec{H}_\Omega^{\vec{J}_\Omega^b}\}$ and current $\vec{J}_\Omega^a/\vec{J}_\Omega^b$ satisfy the following Maxwell's equations:

$$\begin{aligned}\nabla\times\vec{H}_\Omega^{\vec{J}_\Omega^{a/b}}(\vec{r}) &= \vec{J}_\Omega^{a/b}(\vec{r})+j\omega\vec{\vec{\varepsilon}}_\Omega(\vec{r})\cdot\vec{E}_\Omega^{\vec{J}_\Omega^{a/b}}(\vec{r}) \\ \nabla\times\vec{E}_\Omega^{\vec{J}_\Omega^{a/b}}(\vec{r}) &= -j\omega\vec{\vec{\mu}}_\Omega(\vec{r})\cdot\vec{H}_\Omega^{\vec{J}_\Omega^{a/b}}(\vec{r})\end{aligned} \quad \text{(B-8)}$$

for any $\vec{r}\in\Omega$. Inserting $\vec{P}=\vec{E}_\Omega^{\vec{J}_\Omega^a}(\vec{r})$ and $\vec{Q}=\vec{E}_\Omega^{\vec{J}_\Omega^b}(\vec{r})$ and $\vec{\vec{\alpha}}=\vec{\vec{\mu}}_\Omega(\vec{r})$ into (B-7), the following relation can be derived:

$$-j\omega \iiint_\Omega \vec{E}_\Omega^{J_\Omega^a}(\vec{r}) \cdot \vec{J}_\Omega^b(\vec{r}) d\Omega$$
$$+j\omega \iiint_\Omega \vec{J}_\Omega^a(\vec{r}) \cdot \vec{E}_\Omega^{J_\Omega^b}(\vec{r}) d\Omega$$
$$+\omega^2 \iiint_\Omega \vec{E}_\Omega^{J_\Omega^a}(\vec{r}) \cdot \left[ \vec{\varepsilon}_\Omega(\vec{r}) \cdot \vec{E}_\Omega^{J_\Omega^b}(\vec{r}) \right] d\Omega$$
$$-\omega^2 \iiint_\Omega \left[ \vec{\varepsilon}_\Omega(\vec{r}) \cdot \vec{E}_\Omega^{J_\Omega^a}(\vec{r}) \right] \cdot \vec{E}_\Omega^{J_\Omega^b}(\vec{r}) d\Omega \qquad \text{(B-9)}$$
$$= -j\omega \oiint_{\partial\Omega} \hat{n}_{\to\Omega} \cdot \left[ \left\{ \vec{\mu}_\Omega^{-1}(\vec{r}) \cdot \left[ \vec{\mu}_\Omega(\vec{r}) \cdot \vec{H}_\Omega^{J_\Omega^a}(\vec{r}) \right] \right\} \times \vec{E}_\Omega^{J_\Omega^b}(\vec{r}) \right] dS$$
$$- j\omega \oiint_{\partial\Omega} \hat{n}_{\to\Omega} \cdot \left[ \vec{E}_\Omega^{J_\Omega^a}(\vec{r}) \times \left\{ \vec{\mu}_\Omega^{-1}(\vec{r}) \cdot \left[ \vec{\mu}_\Omega(\vec{r}) \cdot \vec{H}_\Omega^{J_\Omega^b}(\vec{r}) \right] \right\} \right] dS$$

Due to the symmetry of $\vec{\varepsilon}_\Omega(\vec{r})$, $\vec{\mu}_\Omega(\vec{r})$, and $\vec{\mu}_\Omega^{-1}(\vec{r})$ and the associative property corresponding to the dot product of two dyads, the following relationship exists:

$$\vec{E}_\Omega^{J_\Omega^a}(\vec{r}) \cdot \left[ \vec{\varepsilon}_\Omega(\vec{r}) \cdot \vec{E}_\Omega^{J_\Omega^b}(\vec{r}) \right] = \left[ \vec{\varepsilon}_\Omega(\vec{r}) \cdot \vec{E}_\Omega^{J_\Omega^a}(\vec{r}) \right] \cdot \vec{E}_\Omega^{J_\Omega^b}(\vec{r}) \quad \text{(B-10.1)}$$
$$\vec{\mu}_\Omega^{-1}(\vec{r}) \cdot \left[ \vec{\mu}_\Omega(\vec{r}) \cdot \vec{H}_\Omega^{J_\Omega^a}(\vec{r}) \right] = \vec{H}_\Omega^{J_\Omega^a}(\vec{r}) \quad \text{(B-10.2)}$$
$$\vec{\mu}_\Omega^{-1}(\vec{r}) \cdot \left[ \vec{\mu}_\Omega(\vec{r}) \cdot \vec{H}_\Omega^{J_\Omega^b}(\vec{r}) \right] = \vec{H}_\Omega^{J_\Omega^b}(\vec{r}). \quad \text{(B-10.3)}$$

Inserting (B-10) into (B-9), (B-9) can be simplified as follows:

$$\iiint_\Omega \vec{E}_\Omega^{J_\Omega^a}(\vec{r}) \cdot \vec{J}_\Omega^b(\vec{r}) d\Omega + \oiint_{\partial\Omega} \vec{E}_\Omega^{J_\Omega^a}(\vec{r}) \cdot \left[ \hat{n}_{\to\Omega} \times \vec{H}_\Omega^{J_\Omega^b}(\vec{r}) \right] dS$$
$$= \iiint_\Omega \vec{J}_\Omega^a(\vec{r}) \cdot \vec{E}_\Omega^{J_\Omega^b}(\vec{r}) d\Omega + \oiint_{\partial\Omega} \left[ \hat{n}_{\to\Omega} \times \vec{H}_\Omega^{J_\Omega^a}(\vec{r}) \right] \cdot \vec{E}_\Omega^{J_\Omega^b}(\vec{r}) dS \qquad \text{(B-11)}$$

Similarly to (A-12), the following boundary currents are defined:

$$\vec{J}_{\partial\Omega}^a(\vec{r}) \triangleq \hat{n}_{\to\Omega}(\vec{r}) \times \left[ \vec{H}_\Omega^{J_\Omega^a}(\vec{r}') \right]_{\vec{r}' \to \vec{r}}, \quad (\vec{r} \in \partial\Omega) \quad \text{(B-12.1)}$$
$$\vec{J}_{\partial\Omega}^b(\vec{r}) \triangleq \hat{n}_{\to\Omega}(\vec{r}) \times \left[ \vec{H}_\Omega^{J_\Omega^b}(\vec{r}') \right]_{\vec{r}' \to \vec{r}}, \quad (\vec{r} \in \partial\Omega). \quad \text{(B-12.2)}$$

Inserting (B-12) into (B-11), (B-11) becomes the following form:

$$\iiint_\Omega \vec{E}_\Omega^{J_\Omega^a}(\vec{r}) \cdot \vec{J}_\Omega^b(\vec{r}) d\Omega + \oiint_{\partial\Omega} \vec{E}_\Omega^{J_\Omega^a}(\vec{r}) \cdot \vec{J}_{\partial\Omega}^b(\vec{r}) dS$$
$$= \iiint_\Omega \vec{J}_\Omega^a(\vec{r}) \cdot \vec{E}_\Omega^{J_\Omega^b}(\vec{r}) d\Omega + \oiint_{\partial\Omega} \vec{J}_{\partial\Omega}^a(\vec{r}) \cdot \vec{E}_\Omega^{J_\Omega^b}(\vec{r}) dS \qquad \text{(B-13)}$$

If the currents in (B-13) are as follows:

$$\left\{ \vec{J}_\Omega^a(\vec{r}), \vec{J}_{\partial\Omega}^a(\vec{r}) \right\} = \left\{ \hat{\xi} \delta(\vec{r} - \vec{r}_a) \right\} \quad \text{(B-14.1)}$$
$$\left\{ \vec{J}_\Omega^b(\vec{r}), \vec{J}_{\partial\Omega}^b(\vec{r}) \right\} = \left\{ \hat{\zeta} \delta(\vec{r} - \vec{r}_b) \right\} \quad \text{(B-14.2)}$$

where $\xi, \zeta = x, y, z$, then

$$\vec{E}_\Omega^{J_\Omega^a}(\vec{r}) = \vec{G}_{\Omega;\xi}^{JE}(\vec{r}, \vec{r}_a)$$
$$= \hat{x} G_{\Omega;x\xi}^{JE}(\vec{r}, \vec{r}_a) + \hat{y} G_{\Omega;y\xi}^{JE}(\vec{r}, \vec{r}_a) + \hat{z} G_{\Omega;z\xi}^{JE}(\vec{r}, \vec{r}_a) \quad \text{(B-15.1)}$$
$$\vec{E}_\Omega^{J_\Omega^b}(\vec{r}) = \vec{G}_{\Omega;\zeta}^{JE}(\vec{r}, \vec{r}_b)$$
$$= \hat{x} G_{\Omega;x\zeta}^{JE}(\vec{r}, \vec{r}_b) + \hat{y} G_{\Omega;y\zeta}^{JE}(\vec{r}, \vec{r}_b) + \hat{z} G_{\Omega;z\zeta}^{JE}(\vec{r}, \vec{r}_b) \quad \text{(B-15.2)}$$

Inserting (B-14) and (B-15) into (B-13), the following relationship is derived:

$$G_{\Omega;\zeta\xi}^{JE}(\vec{r}_b, \vec{r}_a) = \iiint_{\Omega \cup \partial\Omega} \vec{G}_{\Omega;\xi}^{JE}(\vec{r}, \vec{r}_a) \cdot \left[ \hat{\zeta} \delta(\vec{r} - \vec{r}_b) \right] d\Pi$$
$$= \iiint_{\Omega \cup \partial\Omega} \left[ \hat{\xi} \delta(\vec{r} - \vec{r}_a) \right] \cdot \vec{G}_{\Omega;\zeta}^{JE}(\vec{r}, \vec{r}_b) d\Pi \quad \text{(B-16)}$$
$$= G_{\Omega;\xi\zeta}^{JE}(\vec{r}_a, \vec{r}_b)$$

Thus, the following symmetry of Green's function $\vec{\vec{G}}_\Omega^{JE}(\vec{r}, \vec{r}')$ is obtained:

$$\vec{\vec{G}}_\Omega^{JE}(\vec{r}, \vec{r}') = \left[ \vec{\vec{G}}_\Omega^{JE}(\vec{r}', \vec{r}) \right]^T. \quad \text{(B-17)}$$

Similarly, the following symmetry of Green's function $\vec{\vec{G}}_\Omega^{MH}(\vec{r}, \vec{r}')$ can also be obtained:

$$\vec{\vec{G}}_\Omega^{MH}(\vec{r}, \vec{r}') = \left[ \vec{\vec{G}}_\Omega^{MH}(\vec{r}', \vec{r}) \right]^T. \quad \text{(B-18)}$$

Based on the symmetries (B-6), (B-17), and (B-18), (A-10) and (A-11) can be alternatively written as follows:

$$\vec{E}_\Omega(\vec{r}) = \iiint_\Omega \vec{\vec{G}}_\Omega^{JE}(\vec{r}, \vec{r}') \cdot \vec{J}_\Omega(\vec{r}') d\Omega'$$
$$+ \iiint_\Omega \vec{\vec{G}}_\Omega^{ME}(\vec{r}, \vec{r}') \cdot \vec{M}_\Omega(\vec{r}') d\Omega'$$
$$+ \oiint_{\partial\Omega} \vec{\vec{G}}_\Omega^{JE}(\vec{r}, \vec{r}') \cdot \left[ \hat{n}_{\to\Omega} \times \vec{H}_\Omega(\vec{r}') \right] dS' \quad \text{(A-10')}$$
$$+ \oiint_{\partial\Omega} \vec{\vec{G}}_\Omega^{ME}(\vec{r}, \vec{r}') \cdot \left[ \vec{E}_\Omega(\vec{r}') \times \hat{n}_{\to\Omega} \right] dS'$$

$$\vec{H}_\Omega(\vec{r}) = \iiint_\Omega \vec{\vec{G}}_\Omega^{JH}(\vec{r}, \vec{r}') \cdot \vec{J}_\Omega(\vec{r}') d\Omega'$$
$$+ \iiint_\Omega \vec{\vec{G}}_\Omega^{MH}(\vec{r}, \vec{r}') \cdot \vec{M}_\Omega(\vec{r}') d\Omega'$$
$$+ \oiint_{\partial\Omega} \vec{\vec{G}}_\Omega^{JH}(\vec{r}, \vec{r}') \cdot \left[ \hat{n}_{\to\Omega} \times \vec{H}_\Omega(\vec{r}') \right] dS' \quad \text{(A-11')}$$
$$+ \oiint_{\partial\Omega} \vec{\vec{G}}_\Omega^{MH}(\vec{r}, \vec{r}') \cdot \left[ \vec{E}_\Omega(\vec{r}') \times \hat{n}_{\to\Omega} \right] dS'$$

for any $\vec{r} \in \Omega$. In fact, the above (A-10') and (A-11') can be uniformly written as follows:

$$\vec{F}_\Omega(\vec{r}) = \iiint_\Omega \vec{\vec{G}}_\Omega^{JF}(\vec{r}, \vec{r}') \cdot \vec{J}_\Omega(\vec{r}') d\Omega'$$
$$+ \iiint_\Omega \vec{\vec{G}}_\Omega^{MF}(\vec{r}, \vec{r}') \cdot \vec{M}_\Omega(\vec{r}') d\Omega'$$
$$+ \oiint_{\partial\Omega} \vec{\vec{G}}_\Omega^{JF}(\vec{r}, \vec{r}') \cdot \left[ \hat{n}_{\to\Omega} \times \vec{H}_\Omega(\vec{r}') \right] dS' \quad \text{(B-19)}$$
$$+ \oiint_{\partial\Omega} \vec{\vec{G}}_\Omega^{MF}(\vec{r}, \vec{r}') \cdot \left[ \vec{E}_\Omega(\vec{r}') \times \hat{n}_{\to\Omega} \right] dS'$$

for any $\vec{r} \in \Omega$, where $F = E, H$. Following the manner to express convolution integrals in [45], (B-19) can be rewritten as the following (B-19') to compact the integral formulations appeared in this paper.

$$\Omega: \vec{F}_\Omega = \left[ \vec{\vec{G}}_\Omega^{JF} * \vec{J}_\Omega \right]_\Omega + \left[ \vec{\vec{G}}_\Omega^{JF} * \left( \hat{n}_{\to\Omega} \times \vec{H}_\Omega \right) \right]_{\partial\Omega}$$
$$+ \left[ \vec{\vec{G}}_\Omega^{MF} * \vec{M}_\Omega \right]_\Omega + \left[ \vec{\vec{G}}_\Omega^{MF} * \left( \vec{E}_\Omega \times \hat{n}_{\to\Omega} \right) \right]_{\partial\Omega}$$
$$= \left[ \vec{\vec{G}}_\Omega^{JF} * \vec{J}_\Omega + \vec{\vec{G}}_\Omega^{MF} * \vec{M}_\Omega \right]_\Omega \qquad \text{(B-19')}$$
$$+ \left[ \vec{\vec{G}}_\Omega^{JF} * \left( \hat{n}_{\to\Omega} \times \vec{H}_\Omega \right) + \vec{\vec{G}}_\Omega^{MF} * \left( \vec{E}_\Omega \times \hat{n}_{\to\Omega} \right) \right]_{\partial\Omega}$$

for any field point in $\Omega$, where $F = E, H$.

## APPENDIX C: INTEGRAL EXPRESSIONS OF THE FIELDS RELATED TO A SIMPLY CONNECTED INHOMOGENEOUS ANISOTROPIC LOSSY MATERIAL BODY $V_{sim}^{mat}$

In this Appendix C, a simply connected inhomogeneous anisotropic lossy material body $V_{sim}^{mat}$, which is placed in inhomogeneous anisotropic lossy environment and excited by incident field $\vec{F}^{inc}$, is considered. The material boundary $\partial V_{sim}^{mat}$ divides the whole Euclidean space $\mathbb{R}^3$ into two parts, the interior of $V_{sim}^{mat}$ (denoted as $\text{int} V_{sim}^{mat}$) and the exterior of $V_{sim}^{mat}$ (denoted as $\text{ext} V_{sim}^{mat}$) as shown in Fig. 4, and they are open sets obviously [38]. When the polarization electric current and magnetization magnetic current models are employed to depict the polarization and magnetization phenomena of material body, there doesn't exist any scattering surface current on material boundary [1], and the scattering volume polarization electric current and the scattering volume magnetization magnetic current on material body are denoted as $\vec{J}_{mat,pol}^{SV}$ and $\vec{M}_{mat,mag}^{SV}$ respectively, where the superscript "$SV$" is the acronyms of term "scattering volume". The scattering volume ohmic electric current is denoted as $\vec{J}_{mat,ohm}^{SV}$. To simplify the symbolic system of this paper, the summation of $\vec{J}_{mat,pol}^{SV}$ and $\vec{J}_{mat,ohm}^{SV}$ is denoted as $\vec{J}^{SV}$, i.e., $\vec{J}^{SV} \triangleq \vec{J}_{mat,pol}^{SV} + \vec{J}_{mat,ohm}^{SV}$; the $\vec{M}_{mat,mag}^{SV}$ is simply denoted as $\vec{M}^{SV}$, i.e., $\vec{M}^{SV} \triangleq \vec{M}_{mat,mag}^{SV}$. In this paper, it is restricted that the currents $\{\vec{J}^{inc}, \vec{M}^{inc}\}$, which lead to the incident field $\vec{F}^{inc}$, distribute on domain $\text{ext} V_{sim}^{mat}$, i.e., the $\{\vec{J}^{inc}, \vec{M}^{inc}\}$ don't distribute on material body.

The incident field $\vec{F}^{inc}$ satisfies Maxwell's equations

$$\begin{aligned} \nabla \times \vec{H}^{inc}(\vec{r}) &= \vec{J}^{inc}(\vec{r}) + j\omega \vec{\varepsilon}_{env;c}(\vec{r}) \cdot \vec{E}^{inc}(\vec{r}) \\ \nabla \times \vec{E}^{inc}(\vec{r}) &= -\vec{M}^{inc}(\vec{r}) - j\omega \vec{\mu}_{env}(\vec{r}) \cdot \vec{H}^{inc}(\vec{r}) \end{aligned} \quad \text{(C-1)}$$

for any $\vec{r} \in \mathbb{R}^3$. In (C-1), $\vec{\varepsilon}_{env;c} \triangleq \vec{\varepsilon}_{env} + (1/j\omega)\vec{\sigma}_{env}$; $\vec{\varepsilon}_{env}$, $\vec{\mu}_{env}$, and $\vec{\sigma}_{env}$ are the permittivity, permeability, and conductivity of environment. The scattering field $\vec{F}^{sca}$ satisfies the following Maxwell's equations:

$$\begin{aligned} \nabla \times \vec{H}^{sca}(\vec{r}) &= \vec{J}^{SV}(\vec{r}) + j\omega \vec{\varepsilon}_{env;c}(\vec{r}) \cdot \vec{E}^{sca}(\vec{r}) \\ \nabla \times \vec{E}^{sca}(\vec{r}) &= -\vec{M}^{SV}(\vec{r}) - j\omega \vec{\mu}_{env}(\vec{r}) \cdot \vec{H}^{sca}(\vec{r}) \end{aligned} \quad \text{(C-2)}$$

for any $\vec{r} \in \mathbb{R}^3$. In (C-2), $\vec{J}^{SV} = j\omega \Delta \vec{\varepsilon}_c \cdot \vec{E}^{tot}$, and $\vec{M}^{SV} = j\omega \Delta \vec{\mu} \cdot \vec{H}^{tot}$; $\Delta \vec{\varepsilon}_c \triangleq \vec{\varepsilon}_c - \vec{\varepsilon}_{env;c}$, and $\Delta \vec{\mu} \triangleq \vec{\mu} - \vec{\mu}_{env}$; $\vec{\varepsilon}_c \triangleq \vec{\varepsilon} + (1/j\omega)\vec{\sigma}$; $\vec{\varepsilon}$, $\vec{\mu}$, and $\vec{\sigma}$ are the permittivity, permeability, and conductivity of material body $V_{sim}^{mat}$. The total field $\vec{F}^{tot}$ on $\text{int} V_{sim}^{mat}$ satisfies the following Maxwell's equations: [1]

$$\begin{aligned} \nabla \times \vec{H}^{tot}(\vec{r}) &= j\omega \vec{\varepsilon}_c(\vec{r}) \cdot \vec{E}^{tot}(\vec{r}) \\ \nabla \times \vec{E}^{tot}(\vec{r}) &= -j\omega \vec{\mu}(\vec{r}) \cdot \vec{H}^{tot}(\vec{r}) \end{aligned} \quad \text{(C-3)}$$

for any $\vec{r} \in \text{int} V_{sim}^{mat}$.

Inserting (C-1) into (B-19'), and letting the $\Omega$ be whole space $\mathbb{R}^3$, and employing the Sommerfeld's radiation condition for the fields and various Green's functions [37], the following integral expression for $\vec{F}^{inc}$ is obtained:

$$\mathbb{R}^3 \; : \; \vec{F}^{inc} = \left[ \ddot{G}_{env}^{JF} * \vec{J}^{inc} + \ddot{G}_{env}^{MF} * \vec{M}^{inc} \right]_{\text{ext} V_{sim}^{mat}} \quad \text{(C-4)}$$

where $F = E, H$, and the subscripts "$env$" on various Green's functions represent that these Green's functions are the environment Green's functions. Inserting (C-1) into (B-19'), and letting the $\Omega$ be $\text{int} V_{sim}^{mat}$, the following integral expression for the $\vec{F}^{inc}$ on $\text{int} V_{sim}^{mat}$ is obtained:

$$\begin{aligned} \text{int} V_{sim}^{mat} \; : \; \vec{F}^{inc} &= \left[ \ddot{G}_{env}^{JF} * (\hat{n}_- \times \vec{H}_-^{inc}) + \ddot{G}_{env}^{MF} * (\vec{E}_-^{inc} \times \hat{n}_-) \right]_{\partial V_{sim}^{mat}} \\ &= \left[ \ddot{G}_{env}^{JF} * (\hat{n}_- \times \vec{H}^{inc}) + \ddot{G}_{env}^{MF} * (\vec{E}^{inc} \times \hat{n}_-) \right]_{\partial V_{sim}^{mat}} \end{aligned} \quad \text{(C-5)}$$

where $F = E, H$, and $\hat{n}_-$ is the inward normal vector of $\partial V_{sim}^{mat}$. The subscript "$-$" in the right-hand side of the first equality in (C-5) is to emphasize that the corresponding fields distribute on the internal surface of $\partial V_{sim}^{mat}$; the second equality in (C-5) is due to the continuity of $\vec{F}^{inc}$ on $\partial V_{sim}^{mat}$, because the source of $\vec{F}^{inc}$ doesn't distribute on $\partial V_{sim}^{mat}$. Inserting (C-1) into (B-19'), and letting the $\Omega$ be $\text{ext} V_{sim}^{mat}$, and employing the radiation condition, the following integral expression for the $\vec{F}^{inc}$ on $\text{ext} V_{sim}^{mat}$ is obtained:

$$\begin{aligned} \text{ext} V_{sim}^{mat} \; : \; \vec{F}^{inc} &= \left[ \ddot{G}_{env}^{JF} * \vec{J}^{inc} + \ddot{G}_{env}^{MF} * \vec{M}^{inc} \right]_{\text{ext} V_{sim}^{mat}} \\ &+ \left[ \ddot{G}_{env}^{JF} * (\hat{n}_+ \times \vec{H}^{inc}) + \ddot{G}_{env}^{MF} * (\vec{E}^{inc} \times \hat{n}_+) \right]_{\partial V_{sim}^{mat}} \end{aligned} \quad \text{(C-6)}$$

where $F = E, H$, and $\hat{n}_+$ is the outward normal vector of $\partial V_{sim}^{mat}$. Comparing (C-4) with (C-6), it can be derived that

$$\begin{aligned} \text{ext} V_{sim}^{mat} \; : \; 0 &= \left[ \ddot{G}_{env}^{JF} * (\hat{n}_+ \times \vec{H}^{inc}) + \ddot{G}_{env}^{MF} * (\vec{E}^{inc} \times \hat{n}_+) \right]_{\partial V_{sim}^{mat}} \\ &= -\left[ \ddot{G}_{env}^{JF} * (\hat{n}_- \times \vec{H}^{inc}) + \ddot{G}_{env}^{MF} * (\vec{E}^{inc} \times \hat{n}_-) \right]_{\partial V_{sim}^{mat}} \end{aligned} \quad \text{(C-7)}$$

where the second equality is due to that $\hat{n}_- = -\hat{n}_+$ on $\partial V_{sim}^{mat}$.

Similarly to deriving the above (C-4)-(C-7) from (B-19') and (C-1), the following integral formulations corresponding to scattering filed can be derived from (B-19') and (C-2):

$$\mathbb{R}^3 \; : \; \vec{F}^{sca} = \left[ \ddot{G}_{env}^{JF} * \vec{J}^{SV} + \ddot{G}_{env}^{MF} * \vec{M}^{SV} \right]_{\text{int} V_{sim}^{mat}} \quad \text{(C-8)}$$

$$\text{ext} V_{sim}^{mat} \; : \; \vec{F}^{sca} = \left[ \ddot{G}_{env}^{JF} * (\hat{n}_+ \times \vec{H}^{sca}) + \ddot{G}_{env}^{MF} * (\vec{E}^{sca} \times \hat{n}_+) \right]_{\partial V_{sim}^{mat}} \quad \text{(C-9)}$$

$$\text{int} V_{sim}^{mat} \; : \; 0 = \left[ \ddot{G}_{env}^{JF} * (\hat{n}_+ \times \vec{H}^{sca}) + \ddot{G}_{env}^{MF} * (\vec{E}^{sca} \times \hat{n}_+) \right]_{\partial V_{sim}^{mat}} \quad \text{(C-10)}$$

where $F = E, H$. In the process to derive (C-9) and (C-10), the conclusion that the tangential components of $\vec{F}^{sca}$ are continuous on $\partial V_{sim}^{mat}$ has been utilized, and this conclusion is based on that there is not scattering surface current on $\partial V_{sim}^{mat}$ [1].

In addition, the following integral formulation corresponding to the total filed $\vec{F}^{tot}$ on $\text{int} V_{sim}^{mat}$ can be derived from (B-19') and (C-3):

$$\text{int} V_{sim}^{mat} \;:\; \vec{F}^{tot} = \left[ \ddot{\vec{G}}_{sim}^{JF} * \left( \hat{n}_{-} \times \vec{H}^{tot} \right) + \ddot{\vec{G}}_{sim}^{MF} * \left( \vec{E}^{tot} \times \hat{n}_{-} \right) \right]_{\partial V_{sim}^{mat}} \quad \text{(C-11)}$$

where $F = E, H$, and the Green's functions are the ones corresponding to material body $V_{sim}^{mat}$.

ACKNOWLEDGEMENT

This work is dedicated to my mother for her constant understanding, support, and encouragement.

REFERENCES

[1] M. Sancer, K. Sertel, J. Volakis, and P. Alstine, "On volume integral equations," *IEEE Trans. Antennas Propag.*, vol. 54, no. 5, pp. 1488-1495, May 2006.

[2] C. Balanis, *Advanced Engineering Electromagnetics*, 2nd ed. New York, USA: Wiley, 2012.

[3] C. Huygens, *Traite de la Lumiere*, Leyeden, 1690. Translated into English by S. P. Thompson, London, 1912 and reprinted by the University of Chicago Press.

[4] J. Kraus and K. Carver, *Electromagnetics*, 2nd ed. New York, USA: McGraw-Hill, 1973.

[5] T. Young, *A Course of Lectures on Natural Philosophy and Mechanical Arts*. London, UK: 1807.

[6] G. Malyuzhinets, "Developments in Our Concepts of Diffraction Phenomena," *Soviet Physics: Uspekhi*, vol. 69(2), no. 5, pp. 749–758, 1959.

[7] A. Fresnel (1819), "Memoir of the diffraction of light." in *The Wave Theory of Light – Memoirs by Huygens, Young and Fresnel*, H. Crew. New York, USA: American Book Company, 1900.

[8] G. Maggi, "Sulla propagazione libera e perturbata delle onde luminose in un mezzo isotropo," Ann. di Mat. IIa, vol. 16, pp. 21-48, 1888.

[9] A. Rubinowicz, "Die Beugungswelle in der Kirchhoffschen Theorie der Beugungserscheinungen," *Ann. Phys.* vol. 73, pp. 257-278, 1917.

[10] A. Rubinowicz, "Zur Kirchhoffschen Beugungstheories,," *Ann. Phys.* vol. 73, pp. 339-364, 1924.

[11] K. Miyamoto and E. Wolf, "Generalization of Maggi–Rubinowicz theory of the boundary diffraction wave. I, II," *J. Opt. Soc. Am.*, vol. 52, pp. 615–625, 626–637, 1962.

[12] H. v. Helmholtz, *Treatise on Physiological Optics* (translated from German by J. P. C. Southall). New York, USA: Dover, 1962. (Original work published in 1867).

[13] G. Kirchhoff, *Gesammelte Abhandlungen*, Leipzig: Barth, 1882.

[14] G. Kirchhoff, "Zur theorie der lichtstrahlen," reprinted in *Ann. Phys.*, vol. 18, pp. 663-695, 1883.

[15] Lord Rayleigh, "On the passage of waves through apertures in plane screens, and allied problems," *Philos. Mag.*, vol. 43, pp. 259-272, 1897.

[16] A. Sommerfeld, *Optics*. New York, USA: Academic Press, 1954.

[17] K. Zhang and D. Li, *Electromagnetic Theory for Microwaves and Optoelectronics*, 2nd ed. Berlin, Germany: Springer, 2008.

[18] H. Crew, *The Wave Theory of Light – Memoirs by Huygens, Young and Fresnel*. New York, USA: American Book Company, 1900.

[19] M. Born and E. Wolf, *Principles of Optics*, 6th ed. Oxford, UK: Cambridge University Press, 1980.

[20] J. Maxwell, *A Treatise on Electricity and Magnetism*, 2nd ed. London, UK: Clarendon Press, 1881.

[21] A. Love, "The integration of the equations of propagation of electric waves," *Phil. Trans. Roy. Soc. London, Ser. A*, vol. 197, pp. 1–45, 1901.

[22] H. Macdonald, "The diffraction of electric waves round a perfectly reflecting obstacle", *Philosophical Transactions of the Royal Society, Series A*, vol. 210, pp. 113-144, 1911.

[23] S. Schelkunoff, "Some equivalence theorems of electromagnetics and their application to radiation problems", *Bell Syst. Tech. J.*, vol. 15, pp. 92-112, 1936.

[24] J. Larmor, "On the mathematical expression of the principle of Huygens," *Lond. Mathem. Society Proceed.* (2), 1. 1903.

[25] J. Stratton and L. Chu, "Diffraction theory of electromagnetic waves," *Phys. Rev.*, vol. 56, pp. 99–107, 1939.

[26] W. Franz, "Zur Formulierung des Huygensschen Prinzips," *Zeitschrift fur Naturforschung, Band 3a*, pp. 500-506, 1948.

[27] F. Kottler, "Elektromagnetische theorie der Beugung an schwarzen Schirmen," *Ann. Phys.*, vol. 71, pp. 457–508, 1923.

[28] C.-T. Tai, "Kirchhoff theory: Scalar, Vector, or Dyadic?", *IEEE Trans. Antennas Propag.*, vol. 20, no. 1, pp. 114–115, Jan. 1972.

[29] B. Baker and E. T. Copson, *The Mathematical Theory of Huygens' Principle*. Oxford, UK: Clarendon Press, 1939.

[30] R. Collin, *Field Theory of Guided Waves*, 2nd ed. New York, USA: IEEE Press, 1991.

[31] R. Harrington, *Time-Harmonic Electromagnetic Fields*. Hoboken, USA: John Wiley & Sons, 2001.

[32] J. Hadamard, *The Cauchy Problem and the Linear Hyperbolic Partial Differential Equations*. New York, USA: Dover, 1953.

[33] A. Sommerfeld, "Die greensche function der schwingungsgleichung," *Jahresbericht der Deutschen Mathematiker-Vereinigung*, 21, 309-355, 1912.

[34] A. Poggio and E. Miller, "Integral equation solutions of three-dimensional scattering problems," in *Computer Techniques for Electromagnetics*, R. Mittra, Ed., 2nd ed., Hemisphere, New York, 1987.

[35] Y. Chang and R. Harrington, "A surface formulation for characteristic modes of material bodies," *IEEE Trans. Antennas Propag.*, vol. 25, no. 6, pp. 789–795, Nov. 1977.

[36] W. Wu, A. Glisson, and D. Kajfez, "A comparison of two low-frequency formulations for the electric field integral equation," *Tenth Ann. Rev. Prog. Appl. Comput. Electromag.*, vol. 2, pp. 484–491, Monterey, California, Mar. 1994.

[37] C.-T. Tai, *Dyadic Green Functions in Electromagnetic Theory*, 2nd ed. New York, USA: IEEE Press, 1994.

[38] J. Conway, *A Course in Point Set Topology*, Switzerland: Springer, 2014.

[39] D. Miller, "Huygens's wave propagation principle corrected," *Opt. Lett.*, vol. 16, pp. 1370-1372. 1991.

[40] W. Chew, M. Tong, and B. Hu, *Integral Equation Methods for Electromagnetic and Elastic Waves*. London, UK: Morgan & Claypool, 2009.

[41] E. Rothwell and M. Cloud, *Electromagnetics*. Boca Raton, FL: CRC Press, 2001.

[42] C. Oseen, "Uber die wechselwirkung zwischen zwei elektrischen dipolen und uber die drehung der polarisationsebene in kristallen und flussigkeiten," *Annalen der Physik*, vol. 48, pp. 1-56, 1915.

[43] J. Kong, *Electromagnetic Wave Theory*. New York, USA: Wiley, 1986.

[44] W. Chew, *Waves and Fields in Inhomogeneous Media*. New York, USA: IEEE Press, 1995.

[45] A. Peterson, S. Ray, and R. Mittra, *Computational Methods for Electromagnetics*. New York, USA: IEEE Press, 1998.

[46] R. Lian (Dec. 2016), "Surface formulations of the electromagnetic-power-based characteristic mode theory for material bodies — Part II," viXra:1612.0352. Available: https://www.researchgate.net/profile/Renzun_Lian/contributions.

[47] R. Garbacz, "Modal expansions for resonance scattering phenomena," *Proc. IEEE*, vol. 53, no. 8, pp. 856–864, Aug. 1965.

[48] R. Harrington and J. Mautz, "Theory of characteristic modes for conducting bodies," *IEEE Trans. Antennas Propag.*, vol. AP-19, no. 5, pp. 622-628, May 1971.

[49] R. Harrington, J. Mautz, and Y. Chang, "Characteristic modes for dielectric and magnetic bodies," *IEEE Trans. Antennas Propag.*, vol. AP-20, no. 2, pp. 194-198, Mar. 1972.

[50] R. Feynman, R. Leighton, and M. Sands, *The Feynman Lectures on Physics*. Reading, MA, USA: Addison-Wesley Publishing Company, 1964.

[51] R. Lian (Oct. 2016), "Surface formulations of the electromagnetic-power-based characteristic mode theory for material bodies — Part I,"

viXra:1610.0345. Available: https://www.researchgate.net/profile/Renzun_Lian/contributions.